\documentclass{aa}  

\usepackage{graphicx}

\usepackage{txfonts}
\usepackage{hyperref}
\hypersetup{
    colorlinks=true,
    linkcolor=cyan,
    citecolor=cyan,
    urlcolor=cyan,
    }
\usepackage{booktabs}
\newcommand{\orcid}[1]{\protect\href{https://orcid.org/#1}{\protect\includegraphics[width=8pt]{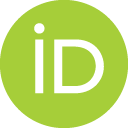}}}

\begin{document}

 \title{Analysis of the \textit{Gaia} DR3 planetary nebula candidates and the possible symbiotic stars among them}

   \author{Lionel Mulato\orcid{0000-0002-6822-9368}\inst{1}
   \and
   Jaroslav Merc\orcid{0000-0001-6355-2468}
          \inst{2,3}
          \and
          Stéphane Charbonnel\orcid{0009-0006-0817-4699}\inst{1}
          \and
          Olivier Garde\orcid{0000-0002-7850-8360}\inst{1}
          \and
          Pascal Le Dû\orcid{0000-0003-2385-0967}\inst{1}
          \and
          Thomas Petit\inst{1}
          }

    \authorrunning{Lionel Mulato et al.}

   \institute{Southern Spectroscopic Project Observatory Team (2SPOT), 45, Chemin du Lac 38690 Châbons, France\\
   \email{lionelmulato@gmail.com}
         \and
             Astronomical Institute of Charles University, V Hole\v{s}ovi\v{c}k{\'a}ch 2, Prague, 18000, Czech Republic
             \\
   \email{jaroslav.merc@mff.cuni.cz}
            \and
            Instituto de Astrof\'isica de Canarias, Calle Vía Láctea, s/n, E-38205 La Laguna, Tenerife, Spain
             }

   \date{Received 7 November 2025;  Accepted 27 April 2026}

 
  \abstract
   {The \textit{Gaia} DR3, released in June 2022, included low-resolution BP/RP (XP) spectra that have been exploited for the classification of various types of emission-line objects using machine-learning techniques. The \textit{Gaia} \emph{Extended Stellar Parametrizer for Emission-Line Stars} (ESP-ELS) algorithm identified 273 sources as potential planetary nebulae (PNe).}
   {We aim to analyze the PN sample produced by the ESP-ELS algorithm to investigate the true nature of the objects classified as PNe.}
   {We extracted all sources from the \textit{Gaia} catalog classified as PNe by the ESP-ELS algorithm and filtered out 200 objects with secure classifications available in the literature. Of these, $\sim$65\% correspond to known Galactic or Magellanic compact PNe, and $\sim$20\% to D- or D'-type symbiotic systems. The XP spectra of the remaining sources were visually inspected, leading to a subset of 14 promising candidates showing strong emission features attributable to H$\alpha$ and [\ion{O}{iii}]~$\lambda$5007. Although typical of PNe, such features are also consistent with D- or D'-type symbiotics, known to mimic compact PNe. We obtained spectroscopic follow-up observations for these objects with the 2SPOT facilities in Chile and France, complemented by an analysis of archival photometric data to further constrain their nature.}
   {We report the identification of nine bona-fide or likely D- or D'-type symbiotic systems, one planetary nebula in the LMC, one polar cataclysmic variable, and three possible Be stars in (or in the direction of) the SMC, within our sample of 14 objects.}
   {}


   \maketitle
%


\section{Introduction}
\label{Introduction}
The advent of large-scale all-sky surveys has transformed the study of emission-line objects by enabling systematic searches based on homogeneous photometric and spectroscopic data for millions of sources. In particular, the third data release of the \textit{Gaia} mission (DR3; \citealt{2016A&A...595A...1G,2023A&A...674A...1G}) provided low-resolution BP/RP (XP) spectra for about 219 million objects, opening new possibilities for the automated identification and classification of emission-line sources.

Within DR3, the \textit{Extended Stellar Parametrizer for Emission-Line Stars} (ESP-ELS; \citealt{2023A&A...674A..26C,2023A&A...674A..28F}) identified and classified 57\,511 emission-line stars (ELSs) into several classes using machine-learning techniques. The classification scheme includes Wolf–Rayet (WR) stars, Be stars, Herbig~Ae/Be stars, T~Tauri stars, active M dwarfs, and planetary nebulae (PNe). While such automated approaches are essential for exploiting the full statistical power of large surveys, their performance depends critically on the ability of low-resolution spectra to distinguish between astrophysical classes with similar observational properties.

PNe are particularly challenging in this respect. Their optical spectra are dominated by strong emission lines, but similar observational characteristics are shared by several other classes of emission-line objects, including compact \ion{H}{ii} regions, some symbiotic stars, cataclysmic variables, Be stars, or WR stars. In the case of the low-resolution \textit{Gaia} XP spectra, many classical diagnostic line ratios cannot be measured, which further increases the degeneracy between these classes. As a result, PN samples derived from automated classifications are expected to contain contaminants. Assessing the reliability and astrophysical content of such samples is therefore essential both for improving PN census studies and for identifying rare or poorly known populations.

The PN candidate sample produced by ESP-ELS provides an opportunity to perform such an assessment. In this work, we analyze the objects classified as PNe in \textit{Gaia} DR3 in order to determine their true nature. Particular attention is given to symbiotic systems, which represent one of the most important classes of PN mimics, especially in the case of dusty systems. This study is part of a broader effort aimed at identifying new interacting binaries and other emission-line objects in \textit{Gaia} DR3 \citep{MercGaia1,2025A&A...695A.227M}.

In the following sections, we first review the main observational properties of PNe and their most common mimics. We then analyse the ESP-ELS PN candidate sample in \textit{Gaia} DR3, identifying correctly classified PNe, contaminants, and previously unrecognized emission-line sources. Selected objects are followed up spectroscopically, and our results are discussed in the context of improving the classification of emission-line objects in large surveys.

\subsection{Spectral diversity of planetary nebulae}
\label{Planetary nebulae: definition and spectral diversity}

Planetary nebulae represent a short ($<10^5$\,yr) evolutionary phase of low- to intermediate-mass stars (1–8\,$M_\odot$), during which material ejected on the asymptotic giant branch is photoionized by the hot white dwarf remnant \citep{2024Galax..12...39K,2022FrASS...9.5287P}. The number of known Galactic PNe has increased significantly in recent decades and currently approaches $\sim$4\,700 objects \citep{2016JPhCS.728c2008P}. However, this remains well below theoretical expectations (6\,600–45\,000), indicating that the Galactic PN census is still highly incomplete \citep{2022FrASS...9.5287P,2006ApJ...650..916M}. Improving the completeness and reliability of PN samples is therefore essential for studies of late stellar evolution and Galactic chemical enrichment.

PNe exhibit a wide range of observational properties across the electromagnetic spectrum, as reviewed, for example, by \citet{2010PASA...27..129F}. Their spectra are typically dominated by recombination and collisionally excited emission lines extending from the ultraviolet (UV) to the near-infrared (NIR). The continuum emission of the white dwarf emerges in the UV, while thermal emission from heated dust ($\sim100$--200\,K) is seen in the mid-infrared (MIR), and bremsstrahlung emission dominates in the radio domain.

The optical appearance of PNe depends strongly on their excitation conditions, evolutionary stage, extinction, and interaction with the interstellar medium \citep{1987A&AS...68...51S,2000oepn.book.....K,2010PASA...27..129F}. For instance, very low-excitation (VLE) PNe are compact objects hosting a relatively cool central star. Their spectra are characterized by prominent H$\alpha$ emission together with [\ion{N}{ii}] $\lambda\lambda$6548, 6583 (signatures of shock excitation), while [\ion{S}{ii}] $\lambda\lambda$6716, 6731 are weak and [\ion{O}{iii}] $\lambda\lambda$4959, 5007 are absent or very faint. In contrast, Type~I PNe exhibit strong [\ion{N}{ii}] $\lambda\lambda$6548, 6583 lines that can exceed H$\alpha$, and [\ion{O}{iii}] $\lambda\lambda$4959, 5007 lines comparable in strength to H$\beta$. The [\ion{O}{iii}] $\lambda$5007/H$\beta$ ratio increases with excitation: low-excitation PNe typically show ratios around $\sim$1, while high-excitation PNe are dominated by strong [\ion{O}{iii}] $\lambda\lambda$4959, 5007 lines and may also display \ion{He}{ii} $\lambda$4686 when the central star is sufficiently hot. In the latter case, [\ion{N}{ii}] $\lambda\lambda$6548, 6583 lines are generally much weaker than H$\alpha$.

Adding to the broad diversity of spectral characteristics exhibited by PNe, a small subset of confirmed PNe host peculiar central stars characterized by fast stellar winds of ionized heavy elements, analogous to those observed in massive WR stars. These CSPNe are denoted as [WR] stars, with brackets distinguishing them from their massive counterparts. The spectra of these PNe display the typical high-excitation nebular lines together with broad stellar emission features of highly ionized helium, nitrogen, carbon, and oxygen, and sometimes a stellar continuum \citep{2003A&A...403..659A,2010PASA...27..129F,2020A&A...640A..10W}. Similar to the massive WR stars, the low mass [WR] stars are classified into three types: [WC] types have prominent helium, carbon, and oxygen lines, [WO] types, whose spectra are similar to that of the WC type but with stronger oxygen lines, and [WN] types are dominated by helium and nitrogen lines. Of the $\sim$130\ [WR] stars catalogued in \cite{2020A&A...640A..10W}, almost all belong to the [WC] and [WO] subclasses; [WN] types are particularly rare \citep{2003MNRAS.346..719M, 2010PASA...27..129F, 2020A&A...640A..10W}. The spectra of PNe with [WC] or [WO] central star are dominated by \ion{C}{iii} $\lambda$$\lambda$4650, 5696, \ion{C}{iv} $\lambda$5806, \ion{He}{ii} $\lambda$$\lambda$4686, 5412, \ion{O}{v} $\lambda$5590 and \ion{O}{vi} $\lambda$$\lambda$3822, 5290 \citep{2003A&A...403..659A, 2020A&A...640A..10W}.

\subsection{Symbiotic stars and other PN mimics}
\label{PNe, symbiotic stars and other mimics}

Planetary nebulae can be confused with several classes of emission-line objects, including compact \ion{H}{ii} regions, supernova remnants, active galaxies, and reddened Be stars \citep{2010PASA...27..129F}. Among these, the most important contaminants of compact PN samples are symbiotic systems.

Symbiotic stars are long-period interacting binaries composed of an evolved giant and a hot compact companion, usually a white dwarf. Mass lost by the giant is partially accreted by the compact star, producing a rich emission-line spectrum and, in many cases, extended ionized nebulae and circumstellar dust \citep{2009AcA....59..169G,2010MNRAS.402.2075A, 2012BaltA..21....5M,2019arXiv190901389M,2025Galax..13...49M}. Despite their astrophysical importance, only about 300 systems are currently known, indicating that the Galactic population is highly incomplete \citep{2019ApJS..240...21A,2019AN....340..598M,2024NatAs...8.1504M,2025A&A...698A.155L}.

\begin{figure}[t]
   \centering
   \includegraphics [width=\columnwidth] {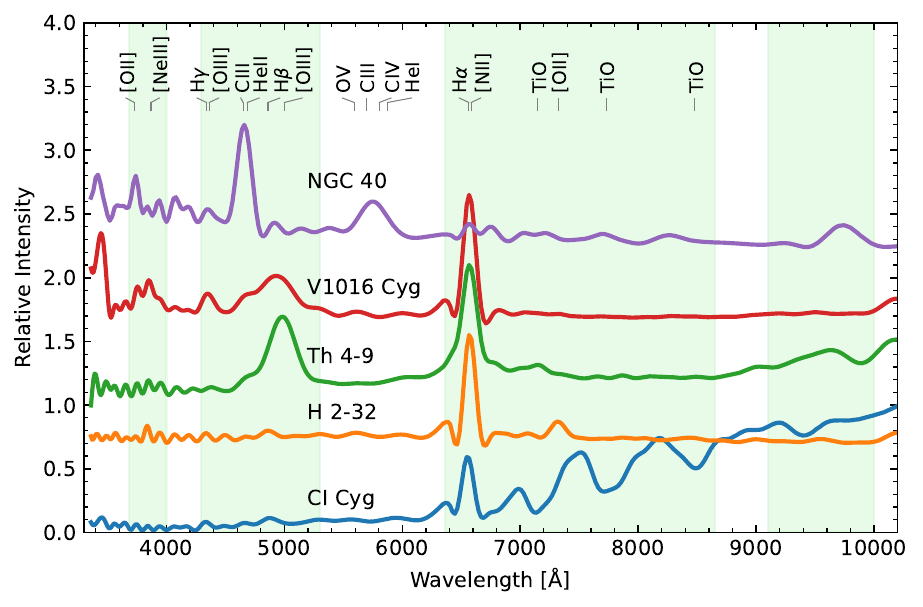}
   \caption{Typical \textit{Gaia} XP spectra of compact PNe and their mimics: NGC~40 ([WR] CSPN), V1016~Cyg (D-type symbiotic star), Th~4-9 (high-excitation PN), H~2-32 (heavily reddened VLE PN), and CI~Cyg (S-type symbiotic star). The green shaded regions show the wavelength ranges used to train the ELS Random Forest classifiers.}
   \label{fig:GAIA_XPSpectra_of_PN}%
\end{figure}

Symbiotic systems are commonly divided into S-, D-, and D'-types based on their near-infrared properties \citep{1982ASSL...95...27A,1995AJ....109.1770M,2019ApJS..240...21A}. S-type systems ("stellar" type) contain a normal late-type giant with little circumstellar dust; their spectra show strong stellar absorption features together with emission lines from the ionized gas around the accretor. Accreting-only S-type systems may show no detectable emission lines, or only low-ionization species such as \ion{H}{i} and \ion{He}{i} \citep[e.g.,][]{2016MNRAS.461L...1M,2019arXiv190901389M,2021MNRAS.505.6121M}, while systems with ongoing shell burning on the white dwarf surface exhibit high-ionization lines such as [\ion{O}{iii}], \ion{He}{ii}, and Raman-scattered \ion{O}{vi} features \citep[][]{2000A&AS..146..407B,2019ApJS..240...21A,2025Galax..13...49M}. These properties generally make S-type systems difficult to confuse with compact PNe of comparable excitation. In some cases, S-type symbiotics host a G- or K-type "yellow" giant instead of an M-type star; such systems often display prominent iron emission lines and are unlikely to be mistaken for PNe, which typically lack a rising red continuum and strong metallic features.

In contrast, D- and D'-type systems ("dusty") can be very challenging to distinguish from PNe. D-types host a pulsating Mira variable, whereas D'-types contain a G- or K-type giant, both embedded in a dense dust shell that obscures the cool component in the optical spectrum. Their spectra may be dominated by strong nebular emission with little or no visible stellar signature, closely resembling those of compact PNe. Infrared diagnostics alone are often insufficient to separate the two classes, since compact PNe may also show emission around $\sim$8\,µm from UV-excited polycyclic aromatic hydrocarbons (PAHs) and around $\sim$20\,µm from dust heated by the central star \citep{2010PASA...27..129F}. The presence of a cool giant can, however, be confirmed through photometric pulsations or near-infrared spectroscopy.

A key physical difference is the higher nebular density in D- and D'-type systems ($\log n_e \sim 6$–>10) compared to typical PNe ($\log n_e \sim 0$–5) \citep{1995PASP..107..462G,2005A&A...435.1087L,2010PASA...27..129F}. As a result, auroral lines such as [\ion{O}{iii}] $\lambda$4363 remain relatively strong compared to the nebular [\ion{O}{iii}] $\lambda\lambda$4959, 5007 lines. Diagnostic diagrams based on ratios such as [\ion{O}{iii}] $\lambda$4363/H$\gamma$ versus [\ion{O}{iii}] $\lambda$5007/H$\beta$ are therefore widely used to distinguish symbiotic systems from PNe \citep{1995PASP..107..462G,2017A&A...606A.110I}. Ratios of [\ion{O}{iii}] $\lambda$4363/H$\gamma \gtrsim 1$ generally indicate a symbiotic nature, although very young and dense PNe may show similar values (e.g., M3-27; \citealt{2024MNRAS.528.4228R}). Because the low spectral resolution of Gaia XP data limits the use of classical diagnostic line ratios, confusion between compact PNe and dusty symbiotic systems is expected in automated classifications.

\section{\textit{Gaia} EPS-ELS workflow, training set, and expected output}
\label{Gaia EPS-ELS and expected output}

The ESP-ELS pipeline is designed to identify and classify emission-line stars (ELS) in \textit{Gaia} DR3 based on BP/RP (XP) spectra. The full workflow is described in the DR3 documentation v1.3\footnote{\hyperlink{https://gea.esac.esa.int/archive/documentation/GDR3/}{https://gea.esac.esa.int/archive/documentation/GDR3/}} \citep{2022gdr3.reptE....V,2022gdr3.reptE..11U}. Sources are first selected as ELS candidates if their H$\alpha$ pseudo-Equivalent Width (pEW) is negative. Only sources brighter than $G = 17.65$ mag and with XP spectra obtained from more than 10 transits with sufficient signal-to-noise ratio are considered.

ESP-ELS uses three sequential classifiers. The first stage assigns a preliminary spectral type tag to all ELS candidates based on their BP/RP spectra. The second classifier, designed specifically to identify PN and WR stars, compares each source with observed BP/RP spectra from the training set. If a source is inconsistent with PNe or WR stars, it is subsequently processed by a third classifier that assigns one of the other ELS types.

The BP/RP spectra cover the wavelength range 336–1020\,nm with a typical resolution $R \sim$ 20–80, depending on wavelength. Because of this low resolution, classical diagnostic lines such as [\ion{O}{iii}]~$\lambda$4363, H$\gamma$, or the [\ion{N}{ii}] doublet are unresolved. As a result, the distinction between compact PNe and D- or D'-type symbiotic systems using the XP spectra alone is inherently limited. Figure~\ref{fig:GAIA_XPSpectra_of_PN} illustrates representative XP spectra of PNe and symbiotic mimics, highlighting the difficulty in separating these classes. The spectra of Th~4-9 (PN) and V~1016~Cyg (D-type symbiotic) appear very similar, whereas the continuum of the late-type giant is evident in CI Cyg spectrum (S-type symbiotic star)\footnote{One should, however, keep in mind that symbiotic stars are strongly variable systems, and even some D-type symbiotics may occasionally develop a pronounced red continuum \citep[see, e.g., the case of RX~Pup;][]{1999MNRAS.305..190M}, thereby resembling S-type systems. This variability further complicates automated classification of symbiotic stars in large surveys.}. The broad stellar \ion{C}{iii} and \ion{C}{iv} emission lines of NGC~40 [WC]-central star are clearly detected but appear blended. 

The PNe set used for training the classifier contains fewer than one hundred objects\footnote{\label{fn:YF}Private communication with Y. Frémat.}, all selected from SIMBAD \citep{2000A&AS..143....9W}. To analyze the composition of the sample, we crossmatched it with the HASH PN database\footnote{\hyperlink{http://202.189.117.101:8999/gpne/index.php}{http://202.189.117.101:8999/gpne/index.php}} (\citealt{2016JPhCS.728c2008P,2017IAUS..323..327B}) and the New Online Database of Symbiotic Variables\footnote{\hyperlink{https://sirrah.troja.mff.cuni.cz/~merc/nodsv/}{https://sirrah.troja.mff.cuni.cz/~merc/nodsv/}}  (NODSV; \citealt{2019RNAAS...3...28M,2019AN....340..598M,Merc+NODSV2025}).

The HASH database contains over 11\,000 entries (as of March 2025), including True (T), Likely (L), and Possible (P) PNe in the Milky Way and the Magellanic Clouds, along with numerous mimics distributed across 40 different classifications. These mimics include objects originally classified as PNe or generic emission-line sources (e.g., by the MASH and IPHAS surveys) but later found to be unrelated to PNe, as well as additional candidates pending verification. The NODSV currently lists approximately 1\,200 entries (as of April 2025), encompassing all confirmed, candidate, and misclassified symbiotic systems in the Milky Way and nearby galaxies.

Our analysis of the PN training set reveals that it is predominantly composed of compact Galactic PNe classified as True ($\sim$70\%) or Likely/Possible ($\sim$13\%). According to the HASH spectra, $\sim$65\% exhibit medium- to high-excitation characteristics, with the remainder showing low to very low excitation. Approximately 40\% host [WR]-type central stars, resulting in an over-representation of this subtype relative to the $\sim$130 [WR] CSPNe currently known \citep{2020A&A...640A..10W}. 

The training set also contains a non-negligible fraction ($\sim$13\%) of symbiotic systems, all classified as D- or D'-type based on the spectra available in HASH.

Consequently, the PN candidate sample provided by ESP-ELS is expected to include compact sources with low- to high-excitation features and point-like morphology ($G < 17.65$ mag). Extended or heavily extincted PNe may be missed, while S-type symbiotic systems, WR stars, Be stars, Herbig Ae/Be stars, T Tauri stars, and active M dwarfs are unlikely to be misclassified as PNe due to their strong continuum. D- and D'-type symbiotic systems, however, can be misclassified as PNe because their emission-line-dominated spectra resemble compact PNe in the limited BP/RP wavelength coverage.

\section{Planetary nebulae in \textit{Gaia} DR3}
\label{PNe in GAIA DR3}

The HASH PN database contains approximately 4\,700 T, L, and P PNe, of which about 3\,000 have a \textit{Gaia} DR3 counterpart. Among these, only 119 were correctly identified as PNe (99 T, 8 L, 12 P), of which 41 are from the training set. In addition, 20 sources were classified as T~Tauri, Herbig~Ae/Be, Be, or WR stars, in three cases where the PN indeed hosts a [WR] central star. As expected (see Section~\ref{Gaia EPS-ELS and expected output}), approximately 80\% of the correctly identified PNe are compact sources, defined either by major diameters smaller than 10\arcsec\ or by a stellar morphology flag in HASH. We also included all Magellanic PNe in this category when this information was unavailable in HASH. Only a small number of objects have angular sizes exceeding 50\arcsec. Overall, Magellanic PNe account for about 35\% of the T, L, P PNe identified by the algorithm.

Of all known T, L, and P PNe, only $\sim$500 sources are compact with $G < 17.65$~mag, thus meeting the criteria for potential correct classification by ESP-ELS. This defines the subset against which the algorithm’s performance should be evaluated. However, only $\sim$20\% of these sources were assigned the PN ClassELS, while the remaining $\sim$80\% lack any classification. Examination of 280 available XP spectra for these unclassified sources shows either spurious spectral features or a lack of detectable H$\alpha$ and/or [\ion{O}{iii}]~$\lambda$5007/H$\beta$ emission.

\begin{figure}[t]
   \centering
   \includegraphics[width=0.49\textwidth]{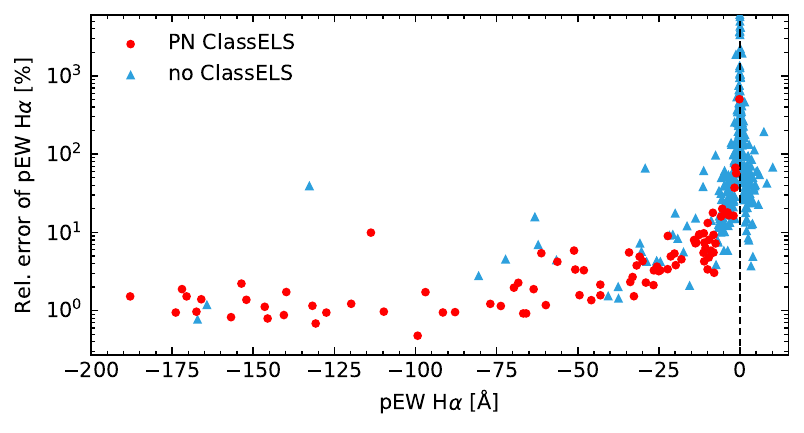}
   \caption{\textit{Gaia} pseudo-equivalent width of the H$\alpha$ line. Compact T, L, P PNe correctly identified by the ESP-ELS algorithm are shown as red circles, while those lacking a ClassELS are shown as blue triangles.}
   \label{fig:PN_ELS_Class}%
\end{figure}

Figure~\ref{fig:PN_ELS_Class} illustrates this behavior using \textit{Gaia} pEW~H$\alpha$ measurements. Sources with a PN ClassELS assignment display strongly negative pEW~H$\alpha$ values with small relative uncertainties (<20\%), confirming their emission-line nature. By contrast, unclassified sources generally exhibit pEW~H$\alpha$ values near zero with large relative errors, reflecting either spurious XP spectra or intrinsically weak emission incompatible with reliable classification.

Interestingly, a subset of $\sim$30 unclassified sources exhibits pEW~H$\alpha$ below $-10$\,\AA\ with relative errors <20\%. Among these, only about half have XP spectra available; for the remaining objects, the data quality was likely insufficient for release. The available spectra show spurious features in the BP range or very weak/absent [\ion{O}{iii}]~$\lambda$5007/H$\beta$, preventing reliable classification.

In summary, the apparent low efficiency of ESP-ELS in identifying known PNe is primarily driven by data-quality limitations. Although PNe are intrinsically strong H$\alpha$ emitters, faint sources or objects with low-quality XP spectra yield unreliable pEW~H$\alpha$ measurements and therefore fail classification. In addition to the pEW~H$\alpha$ criterion, successful classification also requires detectable [\ion{O}{iii}]~$\lambda$5007/H$\beta$ emission and the absence of spurious BP features, which are particularly prone to artifacts.

\section{Analysis of the \textit{Gaia} DR3 PN candidates sample}
\label{Analysis of the GAIA PN candidates sample}

The \textit{Gaia} ESP-ELS algorithm identified 273 sources as potential PNe. Of these, 119 correspond to T, L, or P PNe listed in the HASH PN database (see Section~\ref{PNe in GAIA DR3}). We subsequently examine the remaining objects to identify contaminants and previously unknown emission-line sources of interest.

\subsection{Cross-match with NODSV}
\label{Cross-matching with NODSV}

The \textit{Gaia} DR3 PN candidates sample contains 46 confirmed, likely, possible, or suspected symbiotic stars from the NODSV of which two are extragalactic (in the LMC). Only four objects are classified as S-type symbiotic binaries, while the remainder correspond to D- or D'-type systems, which account for approximately 80\% of all known dusty symbiotic stars. These results are consistent with the expectations discussed in Section~\ref{Gaia EPS-ELS and expected output}.

The 46 NODSV sources overlap with HASH, showing generally good agreement in classification, as 31 are also listed as symbiotic in HASH. Nevertheless, six objects are classified there as likely or probable PNe (Th\,4-11, Mz\,3, Hb\,12, K\,5-37, IRAS\,17272$-$3047, and SMP\,LMC\,63), but not as true PN because they show some possible symbiotic features as noted by the HASH team (except for the last LMC PN). One additional object, CTSS\,2, is classified as a true PN in HASH, although this classification appears questionable. A symbiotic nature of CTSS\,2 was suggested by \citet{2014A&A...567A..12G}. Both the spectrum presented in that study and the one available in HASH show a [\ion{O}{iii}]~$\lambda$4363/H$\gamma$ ratio greater than unity and no detectable optical continuum, suggesting a D- or D'-type nature rather than the classifications proposed in HASH (true PN) and NODSV (possible S-type symbiotic star).

\subsection{Selection of objects of interest}
\label{Selection of objects of interest}
At this stage, we identified 158 objects that were already listed as T, L, or P PNe in HASH or as symbiotic in NODSV. The remaining sources were then cross-matched with SIMBAD, which revealed 14 additional Magellanic PNe not included in HASH. We further removed all objects with a well-established nature (e.g., novae, WR stars, cataclysmic variables, or young stellar objects). This filtering left 74 sources of uncertain or unknown nature, but several of them had promising SIMBAD classifications, such as emission-line stars, long-period variables, Miras, radio or mid-IR sources, and eclipsing binaries.

\begin{table*}[]
    
    \centering
     \small 
     \setlength{\tabcolsep}{4.55pt}
     \renewcommand{\arraystretch}{1.2}
    \caption{List of objects of interest selected from the \textit{Gaia} DR3 PN candidate sample.}
    \begin{tabular}{clcrrrrc}
    \hline\hline
No.&Name$^{a}$ &\textit{Gaia} DR3 & $\alpha_{2000}$ & $\delta_{2000}$ & \textit{G} & pEWH$\alpha$ & SIMBAD \\
& & & [h:m:s] &[d:m:s] &[mag]&[nm] &type$^{b}$\\
    \hline
1&2MASS J00302193-7340258 &4688456119306207104& 00:30:21.92 & -73:40:25.37 &17,39&-4,63&-\\
2&$[$MA93$]$ 332 &4689035080972851072& 00:49:55.93 & -72:24:31.33 &16,98&-2,93&Em*\\
3&2MASS J00520200-7300074 &4685953222889189248& 00:52:01.57 & -73:00:09.76 &17,57&-2,47&RB?\\
4&2MASS J05042350-6946395 &4655200329313892480& 05:04:23.51 & -69:46:39.64 &16,49&-5,54&pA?\\
5&2MASS J10230286-5420071 &5355576935408000256& 10:23:02.86 & -54:20:07.20 &16,93&-9,28&-\\
6&WRAY 15-678 &5338200425455290368& 10:53:24.62 & -60:44:51.79 &16,84&-17,19&Em*\\
7&IRAS 11015-5850 &5338668572545679232& 11:03:38.31 & -59:06:50.61 &14,91&-4,26&MIR\\
8&Mul 22 &4369390734123515392& 17:41:06.38 & -02:25:22.62 &16,2&-11,87&-\\
9&Mul 20 &4369320021783043712& 17:45:06.57 & -02:08:44.21 &16,26&-13,94&-\\
10&Mul 21 &4202825820755995520& 18:59:38.71 & -09:11:03.13 &17,53&-20,43&-\\
11&2MASS J20385125+3111042 &1862451682454046080& 20:38:51.25 & +31:11:04.08 &14,1&-6,19&-\\
12&2MASS J20581257+2920454 &1846427022035230592& 20:58:12.54 & +29:20:44.70 &17,29&-5,25&-\\
13&ATO J315.3668+45.9271 &2163480206474053376& 21:01:28.03 & +45:55:37.74 &14,49&-15,86&LP*\\
14&IRAS 21136+5810 &2191101793978902784& 21:15:01.53 & +58:22:38.77 &14,92&-5,27&*\\

\hline
    \end{tabular}
    \tablefoot{
            $^{a}$ The name corresponds to the SIMBAD identifier when available; otherwise the HASH identifier is used, and when neither is available the 2MASS identifier is given. $^{b}$SIMBAD types correspond to the following classes: pA? - post AGB candidate; RB? - Red Giant Branch star candidate; Em* - emission-line star; MIR - Mid-IR source; LP* - long-period variable; * - star.}
    \label{tab:Objects_of_interest}
\end{table*}

\textit{Gaia} XP spectra were available for 49 of these sources: 35 have unreliable spectra (e.g., with peculiar shapes) or show only a continuum without emission lines, while 14 display prominent emission features around H$\alpha$ and in the [\ion{O}{iii}]~$\lambda$5007/H$\beta$ region. Objects lacking XP spectra, or having unreliable ones, were excluded, as our preliminary multiwavelength analysis (comparison of colors, appearance, etc.) revealed no indication of an emission-line nature (e.g., symbiotic stars, PNe, etc.).

The 14 selected objects of interest are listed in Table~\ref{tab:Objects_of_interest}, and their astrometric parameters are summarized in Table~\ref{tab:GAIA_main}. All of them are unresolved point sources. Among these, three (Mul\,20, Mul\,21, and Mul\,22) are already included in the HASH PN database. These are PN candidates we identified prior to this study using the multiwavelength search techniques described by \citet{2022A&A...666A.152L}. In addition, IRAS\,21136+5810 was also identified as a possible symbiotic star within the general variability classification of \textit{Gaia} DR3 \citep{2023A&A...674A..14R, 2023A&A...674A..13E} and independently identified as a symbiotic candidate by \citet{2025OJAp....8E.122B}. Follow-up of this star was also recently presented by \citet{2025AstBu..80..620T}.

\section{Spectroscopic follow-up}
\label{Spectroscopic follow-up}

We conducted spectroscopic observations of all objects of interest selected in Section~\ref{Selection of objects of interest}, using facilities of the Southern Spectroscopic Project Observatory Team (2SPOT) consortium. Observations were carried out between July 2022 and September 2025 (see Table~\ref{tab:log_obs}). The 2SPOT consortium\footnote{\hyperlink{http://www.2spot.org}{http://www.2spot.org}}, established by amateur astronomers, operates two remotely controlled telescopes in Chile. These include a 0.3-m Ritchey-Chrétien telescope equipped with a medium-resolution Eshel spectrograph ($R = 10\,000$) and a 0.3-m F/4 Newtonian telescope equipped with a low-resolution Alpy600 spectrograph ($R = 600$). For this study, we employed the Alpy600 spectrograph with a 23~$\mu$m slit. Both telescopes are located at Deep Sky Chile (DSC), near Cerro Tololo Observatory.

For northern hemisphere targets, we used our observatories in France. The setups are similar to those in Chile and consist of a 0.2-m F/5 Newtonian or a 0.35-m F/8 Schmidt--Cassegrain telescope, coupled to an Alpy600 or a LISA spectrograph with a 50~$\mu$m slit ($R = 500$). Observations were performed with ATIK 414EX cooled cameras, equipped with 1392\,$\times$\,1040 pixel sensors of 6.45~$\mu$m pixel size. This configuration provides a dispersion of approximately 3.0~\AA/pixel, covering a spectral range of 3800--7800~\AA\ for the Alpy600 and 4000--7500~\AA\ for the LISA.

Typical exposures used 1200~s subframes at 1$\times$1 binning, with total integration times ranging from 60~minutes to several hours, depending on the target brightness. Data reduction was performed using the {\tt Integrated Spectrographic Innovative Software} (ISIS)\footnote{\hyperlink{http://www.astrosurf.com/buil/isis-software.html}{http://www.astrosurf.com/buil/isis-software.html}}. All spectra were corrected for dark, bias, and flat-field frames using standard techniques. The instrumental response was determined from reference spectra of flux standard stars obtained under the same observing conditions as the targets. Wavelength calibration was performed using an argon-neon lamp. No telluric correction was applied to the final spectra.

\section{Results}
\label{Results}

Among our sample of 14 objects, we identify only one source likely to be a new PN in the LMC (Section~\ref{New PN in LMC}). We also report an additional new genuine symbiotic star, which we classified as D-type based on its infrared properties, despite its optical spectrum showing a faint M-type continuum (Section~\ref{New D-type SySt: ATO J315.3668+45.9271}). Eight sources exhibit strong [\ion{O}{iii}]$\lambda$4363 emission, comparable to or stronger than H$\gamma$, suggesting a D- or D'-type symbiotic nature (Section~\ref{New likely D- or D'-types Syst}).  The sample further includes a possible magnetic cataclysmic variable (polar; Section~\ref{New likely polar cataclysmic variable}), and three possible Be stars in the SMC region (Section~\ref{New suspected Be stars in the Magellanic clouds}).

\begin{figure*}[htbp]
  \centering
    \includegraphics[width=0.49\textwidth]{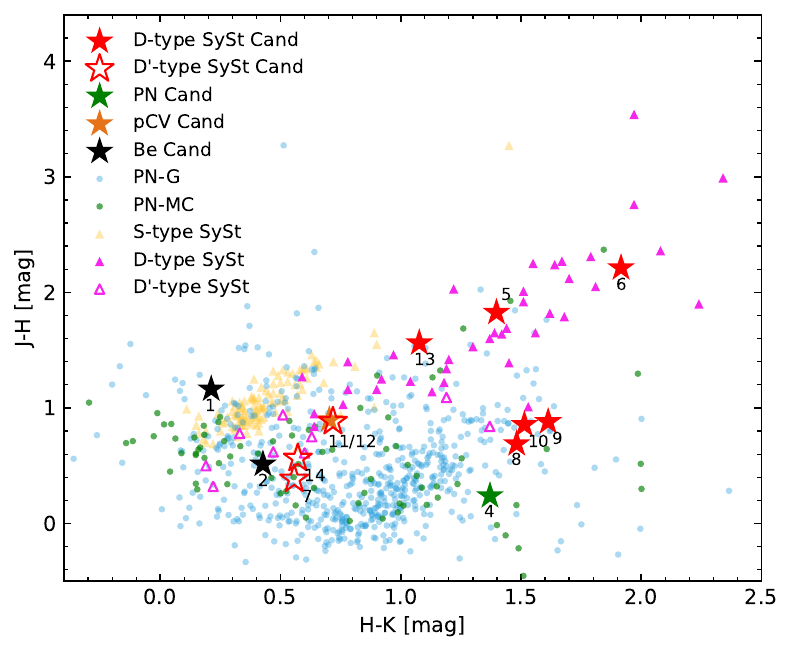}
    \includegraphics[width=0.49\textwidth]{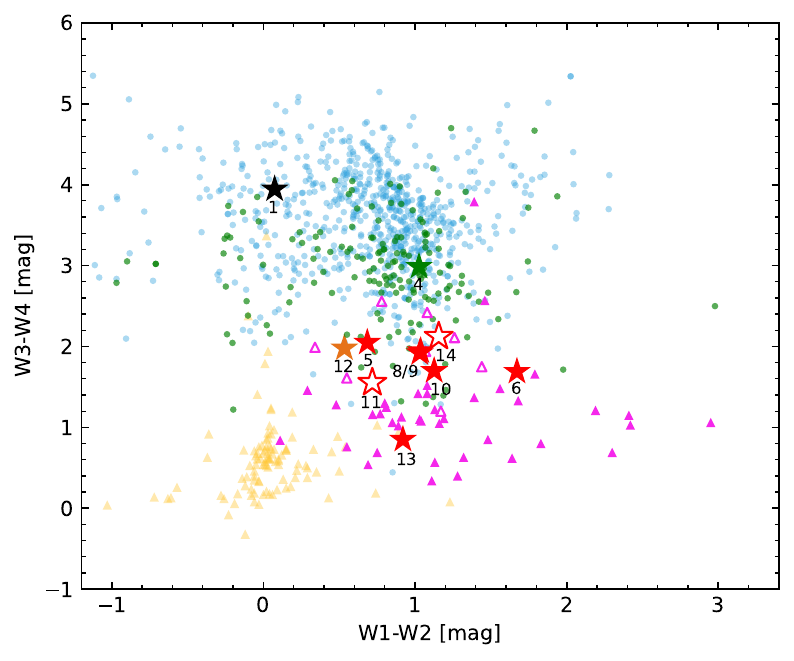}
  \caption{Positions of compact PNe, symbiotic stars, and candidates in the 2MASS (left) and WISE (right) color–color diagrams. 
Known S-type symbiotic systems are shown as yellow triangles, D-type symbiotic stars as filled pink triangles, and D'-type symbiotics as open pink triangles. 
Known compact Galactic PNe are plotted as blue circles, and Magellanic PNe as green circles. 
Objects from this study are represented by star symbols, with colors indicating their suspected nature: D-type symbiotic (filled red), D'-type symbiotic (open red), PNe in the LMC (green), polar CV (orange), and Be stars (black). The numbers correspond to the candidate identifiers listed in Table~\ref{tab:Objects_of_interest}. 
}
  \label{fig:PN_IR_colors}
\end{figure*}

For diagnostic purposes discussed below, Fig.~\ref{fig:PN_IR_colors} shows the positions of our 14 candidates (when data are available) in infrared color–color diagrams constructed from 2MASS \citep{2006AJ....131.1163S} and WISE \citep{2010AJ....140.1868W} observations. The corresponding magnitudes and color indices are provided in Tables~\ref{tab:2MASS_mag} and \ref{tab:WISE_mag}. For comparison, we also plot the locations of compact true PNe from HASH (i.e., PNe with major diameters smaller than 10\arcsec\ or assigned a stellar morphology class) and all confirmed S-, D-, and D'-type symbiotic binaries from NODSV.

\begin{figure}[t]
   \centering
   \includegraphics[width=0.49\textwidth]{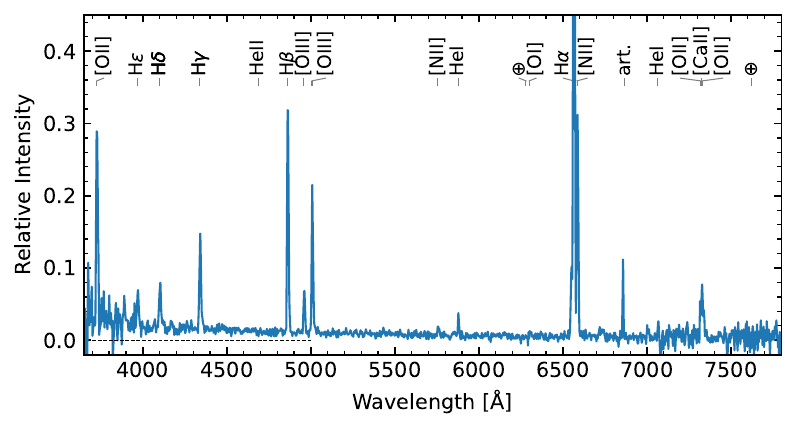}
   \caption{Spectrum of the newly identified PN in the LMC, 2MASS J05042350–6946395, normalized to H$\alpha$.}
   \label{fig:2MASS J05042350-6946395_sp}%
\end{figure}

\subsection{New PN 2MASS J05042350-6946395 in the Large Magellanic Cloud}
\label{New PN in LMC}

2MASS~J05042350-6946395 has been identified as a post-AGB candidate in the LMC \citep{2011A&A...530A..90V}, and more recently suggested to be a pre-PN \citep{2023ApJS..265...18I}.

Our spectrum (see Fig.~\ref{fig:2MASS J05042350-6946395_sp}) shows H$\beta \sim 1.5 \times [\ion{O}{iii}]~\lambda5007$, H$\alpha \sim 3 \times [\ion{N}{ii}]~\lambda6584$, weak \ion{He}{i}~$\lambda5876$, and a slightly rising continuum toward the blue. At first glance, these features are consistent with a PN nature. The Balmer decrement, H$\alpha$/H$\beta \sim 3$, implies relatively low extinction under Case B assumptions, in agreement with the low foreground Galactic extinction in this direction, E(B–V)~$\sim 0.16$~mag \citep{2016ApJ...818..130B}. The observed radial velocity is compatible with LMC membership.

However, the spectrum reveals atypical features for a classical PN. In particular, [\ion{O}{ii}]~$\lambda3727$ is strong, comparable in intensity to H$\beta$ even without extinction correction. We also detect [\ion{O}{ii}]~$\lambda\lambda7319,7330$ and [\ion{Ca}{ii}]~$\lambda7324$, lines commonly associated with fast-shock excitation \citep{1996ApJS..102..161D}, the latter also tracing grain destruction, and frequently observed in Herbig–Haro (HH) objects. Lines that typically dominate HH spectra \citep{1978ApJS...37..117D, 1981ApJS...47..117B, 1991ApJ...376..654G, 1996RMxAA..32..161R}, such as [\ion{S}{ii}]~$\lambda\lambda6716,6731$, are barely detected and remain much fainter than H$\alpha$, while [\ion{O}{i}]~$\lambda\lambda6300,6364$ and [\ion{S}{ii}]~$\lambda\lambda4068,4076$ are absent. This effectively rules out an HH classification, further supported by the lack of nearby star-forming regions.

A weak [\ion{N}{ii}]~$\lambda5755$ line is present despite the modest strength of [\ion{N}{ii}]~$\lambda\lambda6548,6583$, indicating high electron densities. This is further supported by the [\ion{S}{ii}]~6716/6731 ratio of 0.45 reported by \cite{2023ApJS..265...18I}. Its location in IR color–color diagrams (see Fig.~\ref{fig:PN_IR_colors}) is consistent with the distribution of known PNe in the Magellanic Clouds. The SED shows a rising continuum beyond 10~µm, likely due to dust heated by the central star.

We therefore conclude that 2MASS~J05042350-6946395 could be a young, dense PN in the LMC. The presence of hydrogen and helium recombination lines, together with forbidden lines from metals, indicates that the object has evolved beyond its protoplanetary phase, following the definition of \cite{2000oepn.book.....K}.

\subsection{D-type symbiotic star ATO J315.3668+45.9271}
\label{New D-type SySt: ATO J315.3668+45.9271}

\begin{figure}[t]
   \centering
   \includegraphics[width=0.49\textwidth]{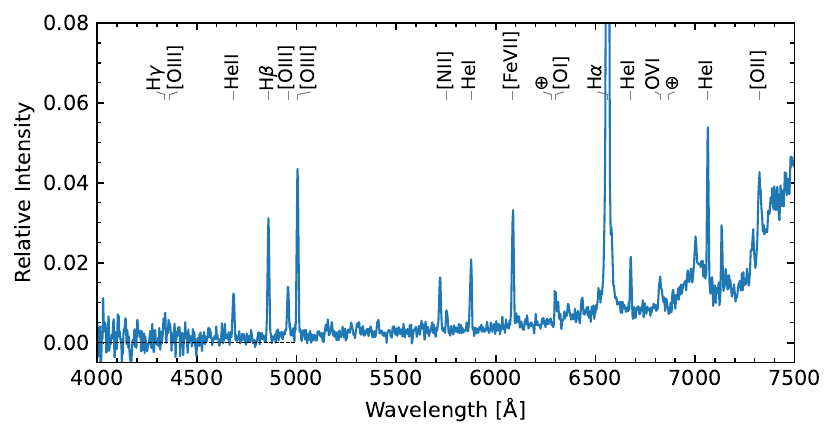}
   \caption{Spectrum of the newly identified D-type symbiotic star, ATO J315.3668+45.9271, normalized to H$\alpha$.}
   \label{fig:ATO J315.3668+45.9271_sp}%
\end{figure}

The spectrum of ATO~J315.3668+45.9271 (see Fig.~\ref{fig:ATO J315.3668+45.9271_sp}) shows a faint M-type continuum together with \ion{H}{i}, \ion{He}{i}, and numerous high-ionization lines, including [\ion{O}{iii}], \ion{He}{ii}, and Raman-scattered \ion{O}{vi}. At first glance, these optical features are consistent with an S-type symbiotic binary. However, the very high H$\alpha$/H$\beta$ ratio ($\sim 32$) indicates substantial extinction. The classification tree of \cite{2019MNRAS.483.5077A} suggests a likely D'-type symbiotic star, although this is not consistent with the spectral type of the cool component.

\begin{figure}[t]
   \centering
   \includegraphics[width=0.49\textwidth]{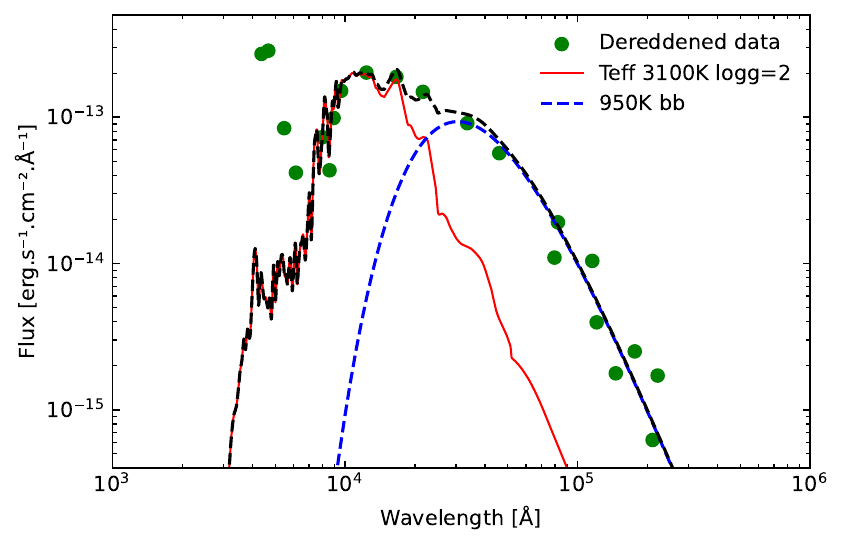}
   \caption{SED of ATO~J315.3668+45.9271 constructed from \textit{Gaia}~DR3, COSMOS, Pan-STARRS, 2MASS, WISE, AKARI, and MSX data. The infrared portion is fitted with a BT-Settl model of a red giant with $T_{\rm eff} = 3100$~K \citep{2014IAUS..299..271A}, combined with a blackbody component at 950~K representing the circumstellar dust.}
   \label{fig:ATO_J315.3668+45.9271_SED}%
\end{figure}

The SED (Fig.~\ref{fig:ATO_J315.3668+45.9271_SED}) peaks near 1~µm and exhibits an infrared excess that cannot be attributed to the giant alone. Fitting the rising profile toward short wavelengths is unreliable due to several unknown parameters, including the luminosity of the hot component, the nebular temperature, and the dominant effect of extinction in this wavelength range. In contrast, IR measurements are less affected by extinction, allowing for more robust modeling. We dereddened the data using the total line-of-sight extinction, $E(B-V) \sim 1.74$, derived from the extinction model of \citet{2016ApJ...818..130B}, which combines the dust maps of \citet{2003A&A...409..205D}, \citet{2006A&A...453..635M}, and \citet{2015ApJ...810...25G}. We then modelled the IR portion of the SED with two components (Fig.~\ref{fig:ATO_J315.3668+45.9271_SED}) at the geometric distance of 3942~pc \citep{2021AJ....161..147B}. The first component is an M-type giant with $T_{\mathrm{eff}} = 3100$~K and $\log(g) = 2$, using BT-Settl models \citep{2014IAUS..299..271A}. The second is a blackbody corresponding to a dust shell at $\sim 950$~K with a radius of $\sim 7.4 \times 10^{13}$~cm, consistent with values of \citet{2007A&A...465..469S} and \citet{2010MNRAS.402.2075A}.

\begin{figure}[t]
   \centering
   \includegraphics[width=0.49\textwidth]{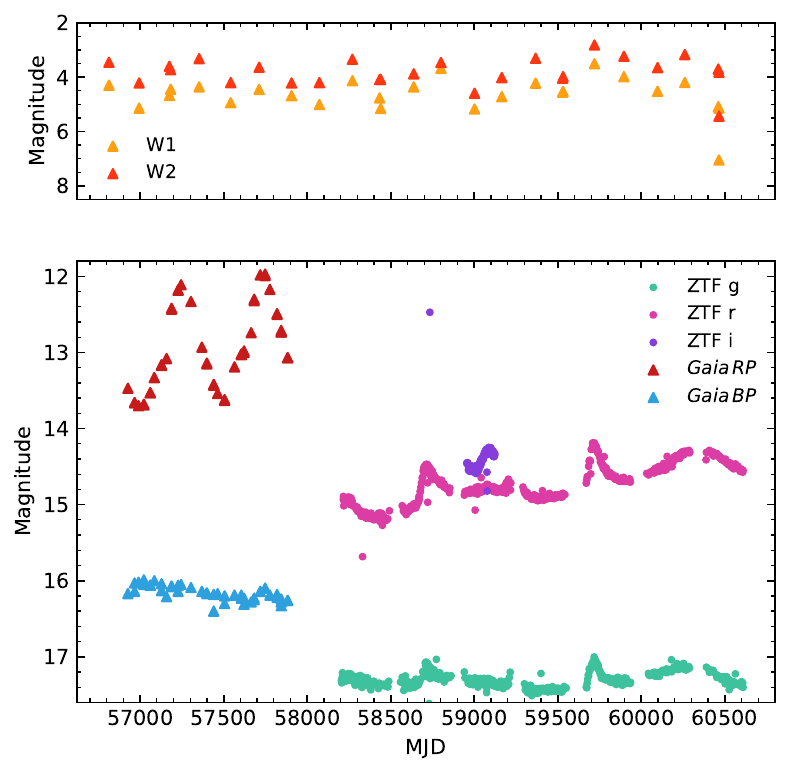}
   \caption{Light curves of ATO J315.3668+45.9271. \textbf{Upper panel}: MIR photometry from WISE. \textbf{Bottom panel}: Optical photometry from \textit{Gaia}, and ZTF surveys. All light curves share the same timescale.}
   \label{fig:ATO_J315.3668+45.9271_var}%
\end{figure}

The light curves (Fig.~\ref{fig:ATO_J315.3668+45.9271_var}) show significant variability in both optical and MIR bands. The WISE $W1$ and $W2$ bands exhibit amplitudes up to $\sim 3$~mag. The $G_{\mathrm{RP}}$ band varies by $\sim 2$~mag with a period of roughly 500~days, consistent with Mira pulsations \citep{1987PASP...99..573W,2009AcA....59..169G,MercGaia1}. In contrast, $G_{\mathrm{BP}}$ shows no comparable periodic variability, instead displaying a gradual decline of $\sim 0.2$~mag over the monitoring period, indicating that the contribution of the giant at these wavelengths is minor. The ZTF $g$ and $r$ bands show smaller variations of $\sim 0.8$~mag, but with a more complex pattern than $G_{\mathrm{RP}}$.

This analysis indicates that ATO~J315.3668+45.9271 is a D-type symbiotic star, with a dust shell not optically thick enough to fully obscure the M-type giant. As mentioned earlier, this star was also classified as a symbiotic candidate in the variability analysis of \textit{Gaia}~DR3 \citep{2023A&A...674A..14R,2023A&A...674A..13E} and by \citet{2025OJAp....8E.122B}, although these works provided no additional information on the star. On the other hand, \citet{2025AstBu..80..620T} obtained spectroscopic follow-up, and their results are fully consistent with our conclusions.

\begin{figure}[t]
   \centering
   \includegraphics [width=0.49\textwidth] {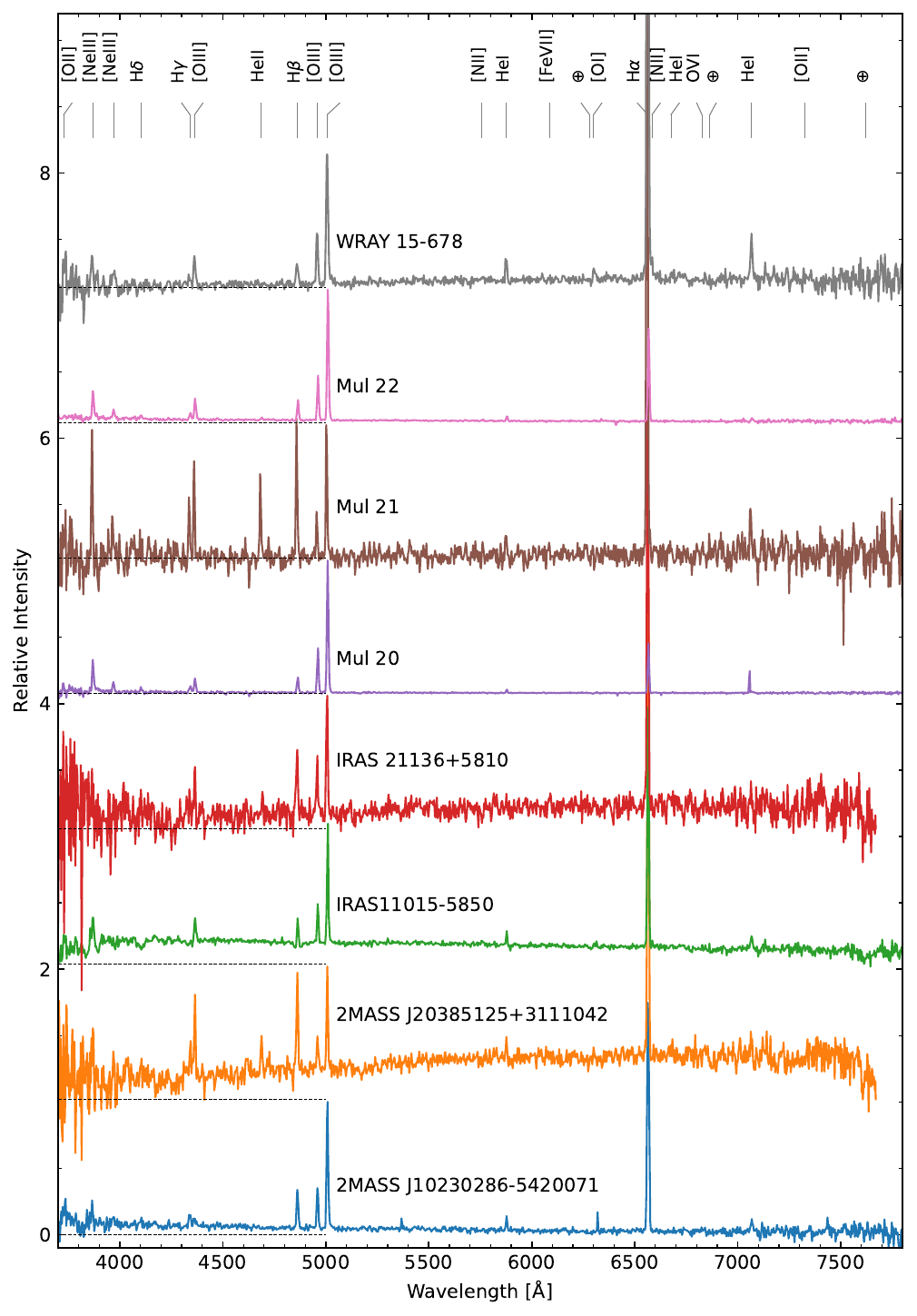}
   \caption{Spectra of the newly identified D'- or D-type symbiotic candidates, showing [\ion{O}{iii}]~$\lambda$4363 stronger than H$\gamma$. The spectra are normalized to [\ion{O}{iii}]~$\lambda$5007. Black dashed lines indicate the zero-intensity level of each spectrum. Telluric absorption bands are marked with the symbol $\oplus$.}
   \label{fig:Dtype_SySt}%
\end{figure}

\begin{figure*}[htbp]
  \centering
    \includegraphics[width=0.49\textwidth]{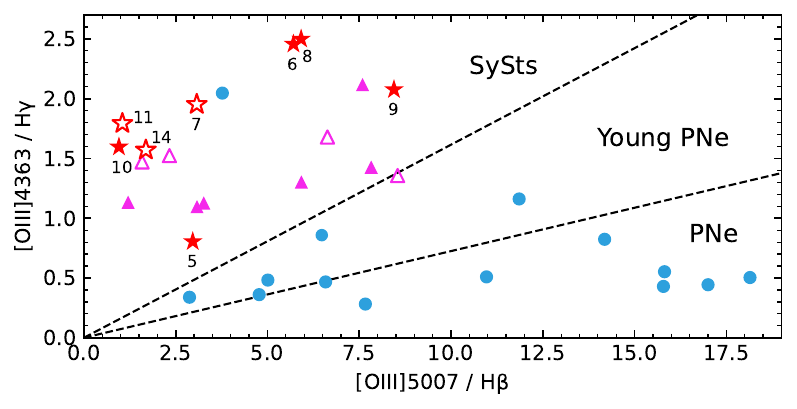}
    \includegraphics[width=0.49\textwidth]{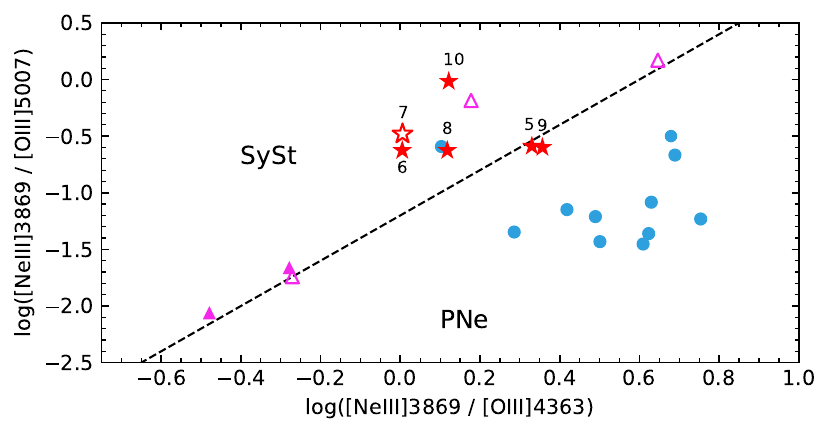}
  \caption{Position of PNe and SySt in the diagnostic diagrams. 
\textbf{Left:} [\ion{O}{iii}] diagram of \cite{1995PASP..107..462G}. 
\textbf{Right:} [\ion{Ne}{iii}]/[\ion{O}{iii}] diagram of \cite{2017A&A...606A.110I}. 
The symbols have the same meaning as in Fig.~\ref{fig:PN_IR_colors}. The numbers correspond to the candidate identifiers listed in Table~\ref{tab:Objects_of_interest}.}
  \label{fig:Gutiérrez-Moreno-Ilkiewicz_diag}
\end{figure*}

\subsection{New D- or D'-type symbiotic candidates with strong [\ion{O}{iii}]$\lambda$4363 emission}
\label{New likely D- or D'-types Syst}

The spectra of the objects discussed in this section are presented in Fig.~\ref{fig:Dtype_SySt}. Mul~20, Mul~22, 2MASS~J10230286-5420071, WRAY~15-678, IRAS~21136+5810, and IRAS~11015-5850 exhibit high-excitation PN-like spectra, with [\ion{O}{iii}]$\lambda$5007 much stronger than H$\beta$ and, in some cases, weak \ion{He}{ii}$\lambda$4686 emission. However, the strength of [\ion{O}{iii}]$\lambda$4363, exceeding that of H$\gamma$, indicates a symbiotic nature. Notably, despite the high-excitation features in WRAY~15-678 suggested by the [\ion{O}{iii}]$\lambda$5007/H$\beta$ ratio, the \ion{He}{ii}$\lambda$4686 line is absent. IRAS~11015-5850 shows a rising blue continuum with prominent Balmer absorption features, which may originate from the cool component, suggesting an F- or G-type giant rather than an M-type giant.

Mul~21 and 2MASS~J20385125+3111042 display spectra inconsistent with classical PNe: the [\ion{O}{iii}] doublet is comparable in strength to both H$\beta$ and \ion{He}{ii}$\lambda$4686, indicating high excitation and electron densities that are atypical for PNe. In both cases, [\ion{O}{iii}]$\lambda$4363 is also very strong, reinforcing a symbiotic classification.

The positions of our eight objects in the 2MASS and WISE color–color diagrams (Fig.~\ref{fig:PN_IR_colors}) are consistent with the distribution of D- or D'-type symbiotics. We propose in Table~\ref{tab:log_obs} a classification of these dusty symbiotic stars based on the analysis of their photometric data. None of the available long-term monitoring surveys, namely the All-Sky Automated Survey for Supernovae \citep[ASAS-SN;][]{2014ApJ...788...48S,2017PASP..129j4502K}, the Asteroid Terrestrial-impact Last Alert System \citep[ATLAS;][]{2018PASP..130f4505T,2020PASP..132h5002S,2021TNSAN...7....1S}, the Zwicky Transient Facility \citep[ZTF;][]{2019PASP..131a8003M}, and \textit{Gaia}~DR3 \citep{2023A&A...674A...1G} in the optical domain, or \textit{NEOWISE} \citep{2011ApJ...731...53M,2014ApJ...792...30M} in the mid-infrared, show evidence of Mira-type pulsations. This lack of variability is most likely due to an optically thick dust shell around the evolved giant \citep{2012BaltA..21....5M}. Only 2MASS~J10230286-5420071 and 2MASS~J20385125+3111042 exhibit maximum amplitude variations of $\sim 0.5$~mag in the WISE $W1$ and $W2$ bands, without any clear periodicity over the 10-year monitoring period. These modest variations may be associated with pulsations of the cool component, but the flux appears to be dominated by dust thermal emission. For 2MASS~J20385125+3111042 and IRAS~21136+5810, ZTF measurements show maximum amplitude variations of $\sim 0.2$~mag in both the $g$ and $r$ bands.

Following the classification scheme of \cite{2019MNRAS.483.5077A}, our eight objects fall among D'-type symbiotic binaries, with the exception of IRAS~11015-5850, whose IR color indices do not allow a clear distinction between the two types. 

We further examined the spectral energy distributions (SEDs) of these sources, shown in Fig.~\ref{fig:SEDs_grid}, collecting the data from Galaxy Evolution Explorer (GALEX) satellite \citep{2014Ap&SS.354..103B}, \textit{Gaia} DR3 \citep[$G_{\mathrm{BP}}$, $G_{\mathrm{RP}}$, and synthetic Johnson \textit{B, V, I,} and Sloan \textit{r, g, i, z};][]{2023A&A...674A...1G}, the Guide Star Catalog II \citep[GSC;][]{2008AJ....136..735L}, the Panoramic Survey Telescope and Rapid Response System \citep[Pan-STARRS;][]{2016arXiv161205560C}, the AAVSO Photometric All-Sky Survey \citep[APASS;][]{2015AAS...22533616H}, the COSMOS survey \citep{2007ApJS..172..196K}, the Two Micron All Sky Survey \citep[2MASS;][]{2006AJ....131.1163S},  VISTA Hemisphere Survey \citep[VHS;][]{2013Msngr.154...35M}, the Deep Near-infrared Southern Sky Survey \citep[DENIS;][]{1997Msngr..87...27E}, the Wide-field Infrared Survey Explorer \citep[WISE;][]{2010AJ....140.1868W}, AKARI \citep{2010A&A...514A...1I}, and Infrared Astronomical Satellite \citep[IRAS;][]{1988iras....1.....B}.

Mul~21, 2MASS~J10230286-5420071, and WRAY~15-678 clearly exhibit a peak intensity between 2 and 4~µm, while IRAS measurements (available only for WRAY~15-678) indicate an additional rise near 10~µm. These are typical IR features of D-type symbiotic stars. The 2–4~µm peak is attributed to the optically thick dust shell obscuring the giant, generally heated to 700–1000~K \citep{1982ASSL...95...27A, 1986AJ.....92.1118K, 1988AJ.....95.1817K, 2019ApJS..240...21A}, whereas the secondary peak near 10~µm originates from a cooler dust shell at approximately 400~K \citep{2010MNRAS.402.2075A, 2018AstL...44..265J}. The positions of 2MASS~J10230286-5420071 and WRAY~15-678 in the 2MASS color–color diagram (Fig.~\ref{fig:PN_IR_colors}) further support a D-type classification rather than D'-type.

In contrast, 2MASS~J20385125+3111042, IRAS~21136+5810, and IRAS~11015-5850 exhibit no 2–4~µm peak and show a declining profile from the optical range. IRAS~11015-5850 displays a faint bump near 8~µm, while the far-IR measurements of IRAS~21136+5810 indicate an increasing flux at 100~µm. These low temperatures, not exceeding $\sim$500~K, point toward a D'-type nature \citep{1984Ap&SS..99..101A, 2007A&A...472..497A}. The positions of IRAS~21136+5810 and IRAS~11015-5850 in the 2MASS color–color diagram (Fig.~\ref{fig:PN_IR_colors}) further support a D'-type classification rather than D-type. The Balmer-line absorption features in the optical spectrum of IRAS~11015-5850 provide additional evidence for its D'-type nature. Finally, the cases of Mul~20 and Mul~22 remain ambiguous due to their relatively flat IR profiles, which suggest a D'-type nature \citep{2019ApJS..240...21A}, but their weak 2–4~µm bumps favor a D-type classification.

Figure~\ref{fig:Gutiérrez-Moreno-Ilkiewicz_diag} shows the positions of these sources in the diagnostic diagrams developed by \cite{1995PASP..107..462G} and \cite{2017A&A...606A.110I} for distinguishing PNe from symbiotic binaries. Line ratios were derived from the maximum intensities of the relevant emission lines (see Table~\ref{tab:lines_intensities}). For comparison, we used spectra available in HASH to plot the positions of known D- or D'-type symbiotics and compact PNe; the reference set is listed in Table~\ref{tab:reference_sp}.
The only known PN falling within the SySt region is M3-27, a well-studied, dense, and very young PN \citep{1997MNRAS.288..777M, 2024MNRAS.528.4228R}. All eight of our objects occupy the D- or D'-type SySt regions of the diagrams, clearly separated from the PN domain.

\begin{figure}[t]
   \centering
   \includegraphics[width=0.49\textwidth]{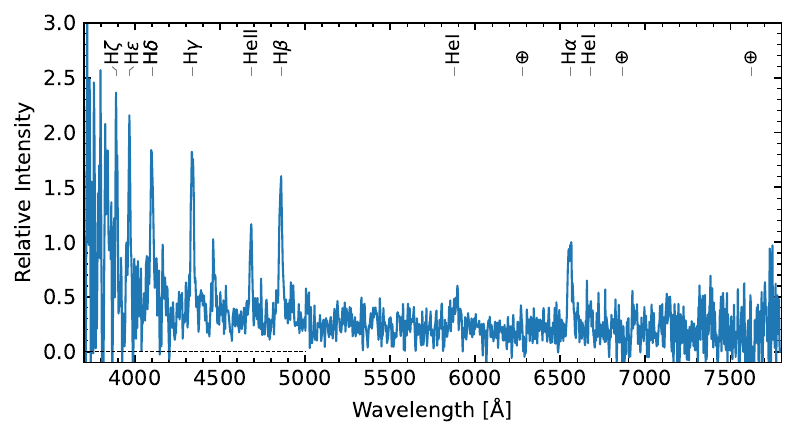}
   \caption{Spectrum of the newly identified polar CV, 2MASS J20581257+2920454, normalized to H$\alpha$.}
   \label{fig:pCV_sp}%
\end{figure}

\begin{figure}[t]
   \centering
   \includegraphics[width=0.49\textwidth]{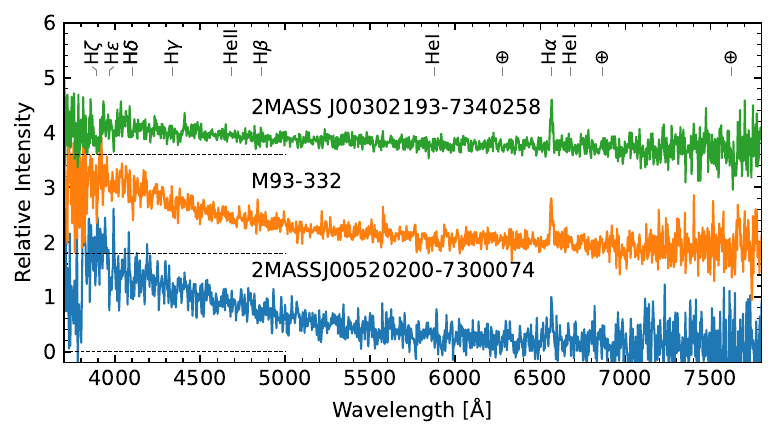}
   \caption{Spectra of the newly identified possible Be stars in the SMC, normalized to H$\alpha$.}
   \label{fig:Be_star}%
\end{figure}

\subsection{New polar 2MASS J00302193-7340258}
\label{New likely polar cataclysmic variable}

The spectrum of 2MASS~J00302193-7340258, shown in Fig.~\ref{fig:pCV_sp}, is clearly not PN-like. It displays broad Balmer emission lines with an inverted decrement, H$\alpha$/H$\beta$~$\approx$~0.62. Weak \ion{He}{i}~$\lambda\lambda$5876,~7065 and strong \ion{He}{ii}~$\lambda$4686 are also present. An X-ray source (1RXS~J205812.8+292037) lies close to the object and was noted by \citet{2006ApJS..163..344K} as a possible counterpart. These spectral characteristics are typical of polar CVs \citep{2017AJ....153..144O, 2020AJ....159..114O, 2023ApJ...957...89O, 2024AJ....167..186S}. The position of the object in the HR diagram ($G_{\rm BP}$--$G_{\rm RP}$~=~0.79~mag, $M_{G}$~=~8.2~mag) is also consistent with the locus of known polar CVs (see Fig.~2 in \citealt{2020MNRAS.492L..40A}). To derive $M_{G}$, we adopted the 3D dust extinction map \citep{2016ApJ...818..130B} and the geometric distance of 659.45~pc from \citet{2021AJ....161..147B}.

\subsection{New possible Be stars in the direction of the Small
 Magellanic Cloud}
\label{New suspected Be stars in the Magellanic clouds}
The spectra of 2MASS~J00302193-7340258, 2MASS~J00520200-7300074, and [MA93]~332 are shown in Fig.~\ref{fig:Be_star}. The first two objects exhibit radial velocities consistent with SMC membership, while the third shows no significant velocity shift, likely due to the low S/N and the fact that the SMC systemic velocity is comparable to our measurement uncertainty ($\sim$100\,km\,s$^{-1}$)\footnote{In the catalog of \citet{2023A&A...672A..65J}, the first object is assigned an almost unit probability of being an SMC member, whereas this probability is lower for the second and nearly negligible for the third, suggesting that the latter two may in fact be foreground stars. A precise radial-velocity measurement would be required to confirm their membership.}. All three spectra display similar characteristics: a hot, stellar-like continuum and weak H$\alpha$ emission. We therefore classify these sources as likely reddened Be stars.

A previous spectroscopic observation of [MA93]~332 was reported by \citet{1993A&AS..102..451M}, who noted the absence of a detectable continuum and only a faint, uncertain H$\alpha$ line. The object was subsequently classified in SIMBAD as an emission-line star. Later, 2MASS~J00520200-7300074 was identified as a red-giant-branch candidate by \citet{2011AJ....142..103B}, based on \textit{Spitzer Space Telescope} infrared observations; however, our spectrum does not support this classification.

\section{Conclusion}
\label{Conclusion}
The spectroscopic coverage provided by \textit{Gaia}~DR3, although in low resolution, combined with the ESP-ELS classification algorithm, offers a new opportunity to explore emission-line objects across the entire sky in a uniform way. In this work, we analyzed the subsample of sources that ESP-ELS classified as potential PNe, with the aim of assessing the reliability of the automated classification and identifying previously unrecognized objects of interest.

The ESP-ELS algorithm applied to the low-resolution XP spectra yielded 273 PN candidates. Among them, 119 correspond to known PNe listed in the HASH database. The remaining sources include a significant number of dusty symbiotic systems catalogued in NODSV ($\sim$80\% of currently known D- and D'-type symbiotics), as expected given their spectroscopic resemblance to compact PNe. 

The relatively small number of genuine PNe recovered by ESP-ELS, compared to the $\sim$4700 catalogued in HASH, can be attributed to several factors. \textit{Gaia} is optimized for point-source detection, and the algorithm imposes a magnitude limit ($G < 17.65$~mag), leaving only $\sim$500 sources potentially suitable for reliable classification. Successful classification further requires that sources are identified as reliable H$\alpha$ emitters, i.e., the H$\alpha$ pseudo-equivalent width measured by ESP-ELS must be negative with low uncertainty. Although PNe are intrinsically strong H$\alpha$ emitters, hundreds of faint objects or those with low-quality XP spectra fail this criterion. Futhermore the classification relies on detectable H$\beta$ and [\ion{O}{iii}]~$\lambda\lambda$4959,5007 emission, which further excludes PNe significantly affected by extinction and exhibiting spurious features in the BP part of the spectrum.

From the ESP-ELS output, we selected 14 objects of particular interest, those with uncertain or unknown classifications in the literature and XP spectra showing prominent emission features. Follow-up spectroscopy was carried out using the 2SPOT facilities in France and Chile. Among these, we identified one PN in the LMC, nine bona fide or likely D- and D'-type symbiotic systems, one magnetic cataclysmic variable (polar), and three possible Be stars in or toward the SMC.

This study illustrates both the potential and the limitations of machine-learning classifications based on \textit{Gaia} XP spectra, as well as of the value of ground-based follow-up observations, particularly in cases where different types of emission-line sources, such as compact PNe and dusty symbiotic systems, exhibit overlapping spectral characteristics (in particular in low resolution). Nevertheless, the availability of a homogeneous, all-sky spectroscopic dataset like that provided by \textit{Gaia}, when combined with targeted ground-based spectroscopy, represents a powerful tool for the discovery, classification, and characterization of diverse classes of emission-line objects. Future releases of \textit{Gaia} data, together with the refinement of training samples and cross-validation with complementary surveys, will further enhance the ability to disentangle these populations.


\begin{acknowledgements}
We thank the referee for their careful reading of the manuscript and constructive comments, which significantly improved the clarity and presentation of this work. The research of J.M. was supported by the Czech Science Foundation (GACR) project no. 24-10608O and by the Spanish Ministry of Science and Innovation with grant no. PID2023-146453NB-100 (PLAtoSOnG).

This work has made use of data from the European Space Agency (ESA) mission {\it Gaia} (\url{https://www.cosmos.esa.int/gaia}), processed by the {\it Gaia} Data Processing and Analysis Consortium (DPAC, \url{https://www.cosmos.esa.int/web/gaia/dpac/consortium}). Funding for the DPAC has been provided by national institutions, in particular the institutions participating in the {\it Gaia} Multilateral Agreement. This work has also made use of the Python package GaiaXPy, developed and maintained by members of the Gaia Data Processing and Analysis Consortium (DPAC), and in particular, Coordination Unit 5 (CU5), and the Data Processing Centre located at the Institute of Astronomy, Cambridge, UK (DPCI). This publication has made use of data products from the Two Micron All Sky Survey, which is a joint project of the University of Massachusetts and the Infrared Processing and Analysis Center/California Institute of Technology, funded by the National Aeronautics and Space Administration and the National Science Foundation.This research has used SIMBAD database \citep{2000A&AS..143....9W}, Vizier catalog access tool \citep{2000A&AS..143...23O}, Aladin Sky Atlas \citep{2000A&AS..143...33B}, and the cross-match service \citep{2012ASPC..461..291B}  provided by CDS, Strasbourg Observatory, France. This research has made use of the SVO Filter Profile Service "Carlos Rodrigo", funded by MCIN/AEI/10.13039/501100011033/ through grant PID2020-112949GB-I00.

\end{acknowledgements}

%
   \bibliographystyle{aa} 
  \bibliography{bibliography.bib} 
%

\begin{appendix}

\begin{figure*}[htbp]
\section{SED of observed objects}\label{appendix}
  \centering
    \includegraphics[width=0.33\textwidth]{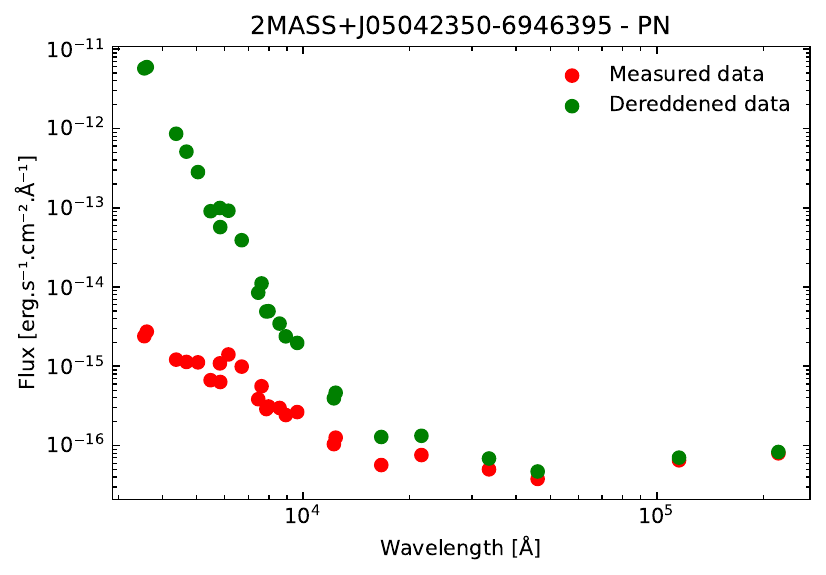}
    \includegraphics[width=0.33\textwidth]{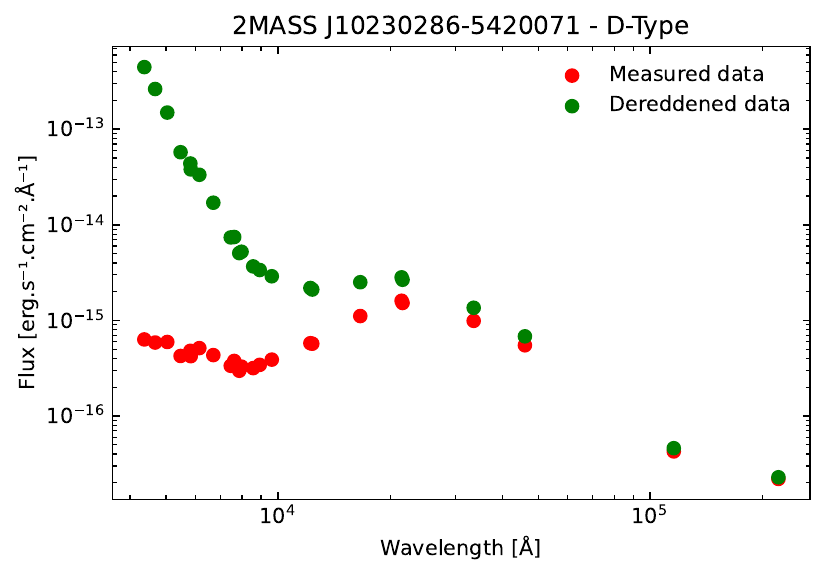}
    \includegraphics[width=0.33\textwidth]{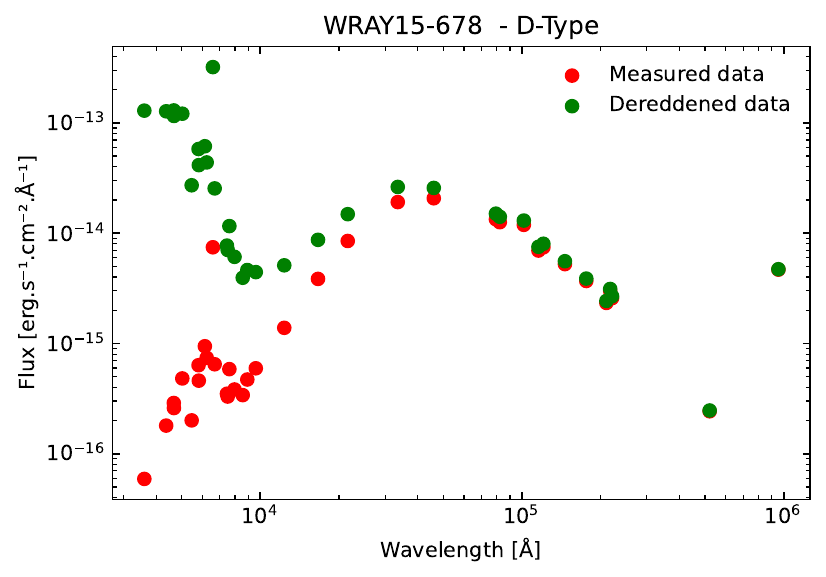}
    \includegraphics[width=0.33\textwidth]{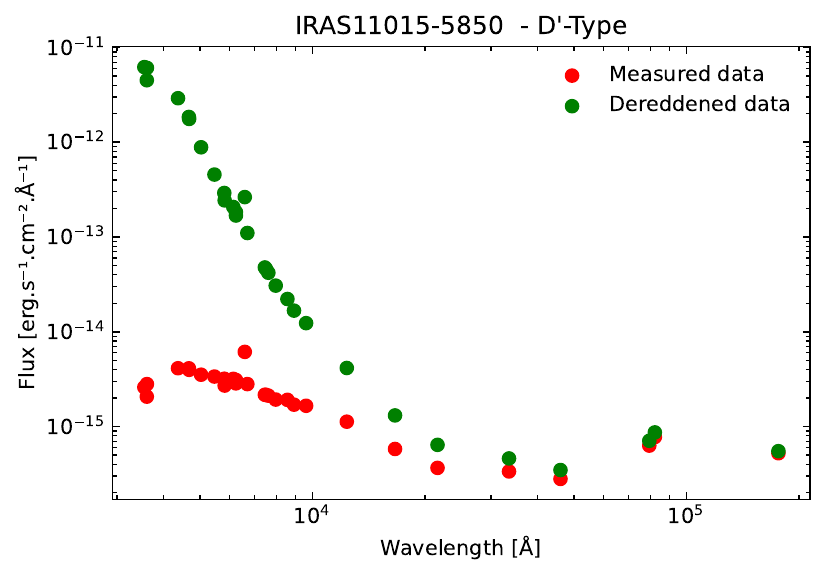}
    \includegraphics[width=0.33\textwidth]{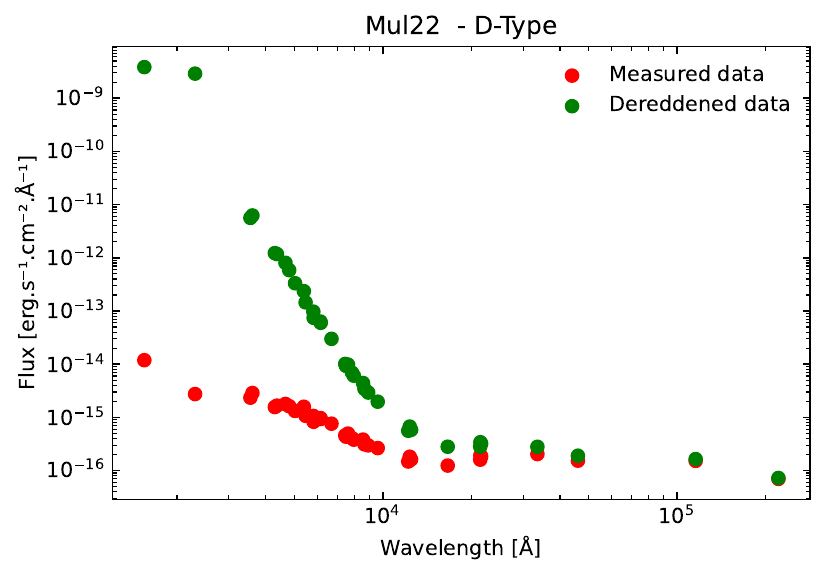}
    \includegraphics[width=0.33\textwidth]{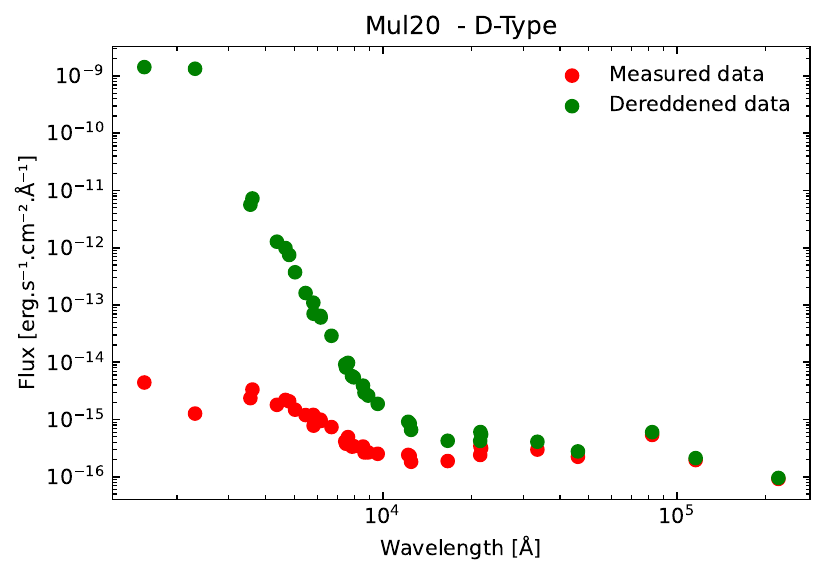}
\includegraphics[width=0.33\textwidth]{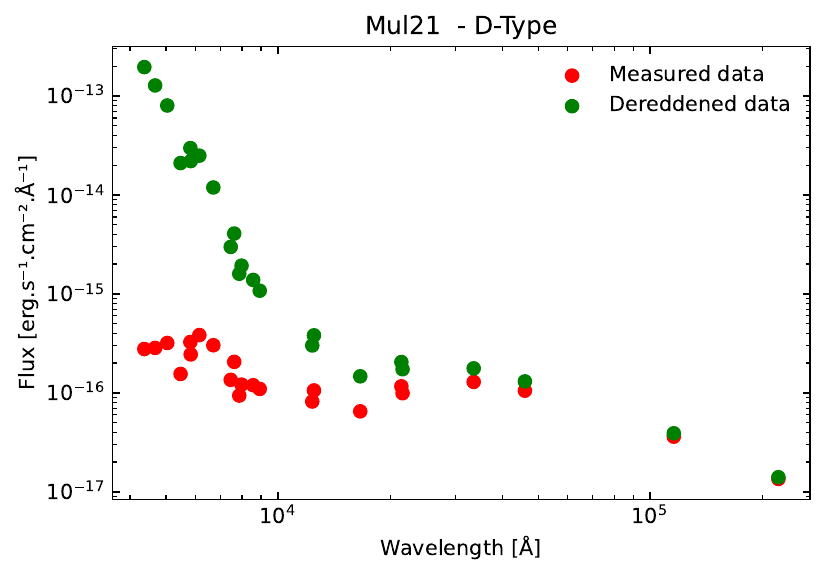}
    \includegraphics[width=0.33\textwidth]{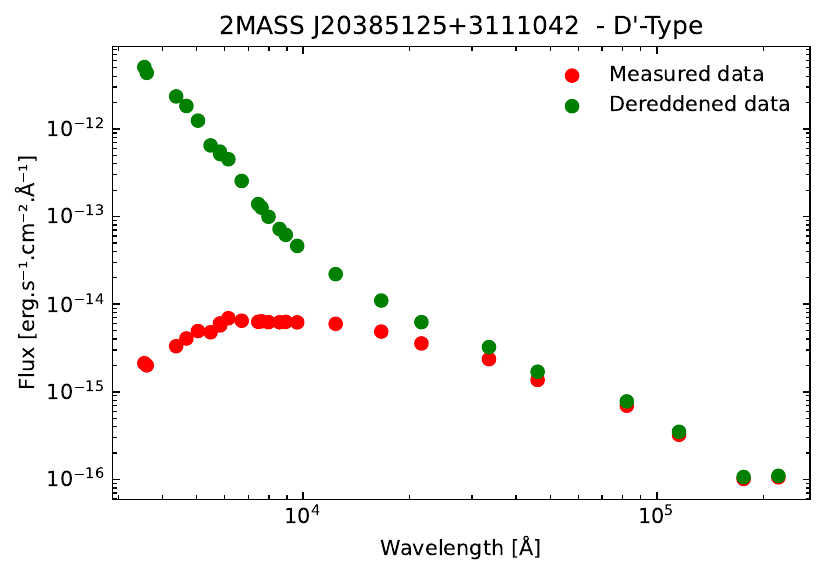}
    \includegraphics[width=0.33\textwidth]{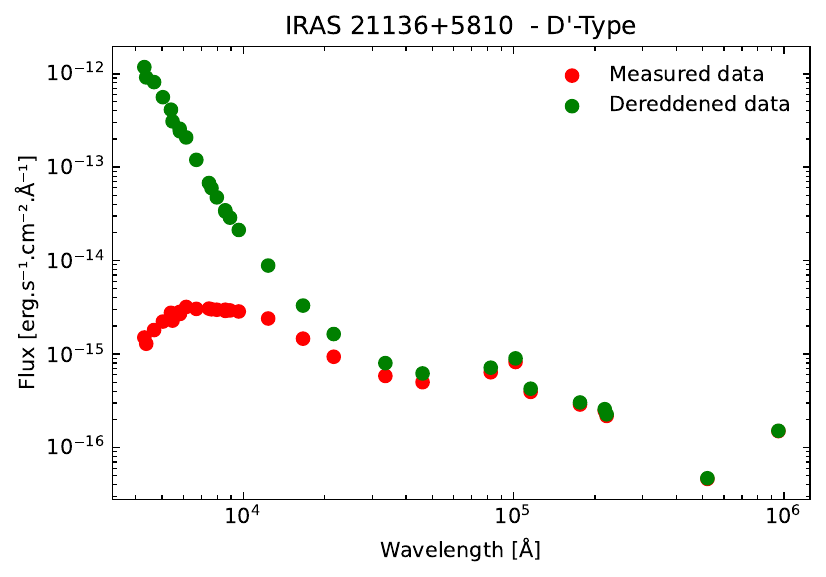}

  \caption{Spectral energy distributions of eight D- or D'-type symbiotic candidates showing [\ion{O}{iii}]~$\lambda$4363 stronger than H$\gamma$, along with the SED of the young, dense PN 2MASS J05042350-6946395. Photometric data were collected from GALEX, \textit{Gaia} DR3, Pan-STARRS, APASS, COSMOS, 2MASS, VHS, DENIS, WISE, AKARI, and IRAS. The dereddened data were computed using the total line-of-sight extinction for each object \citep{2016ApJ...818..130B}.
}
  \label{fig:SEDs_grid}
\end{figure*}

\newpage\onecolumn

\begin{table}
  \section{Additional tables}
    \centering
    \renewcommand{\arraystretch}{1.4}
    \caption{\textit{Gaia} DR3 \citep{2023A&A...674A...1G} parallax $\varpi$, goodness-of-fit (GOF), RUWE of target objects. The geometric distances from \citet{2021AJ....161..147B} are listed and adopted for the calculations in this work. The calculated vertical distances from the Galactic disc are also included in the Table. The last column contains the subclass of each target.}
    \begin{tabular}[t]{lrrrrrrl}
\hline\hline
Object&$\varpi$&$\varpi$/$\sigma_\varpi$&RUWE$^a$&GOF$^b$&Dist. (B-J)&z&Nature\\
&[mas]&&&&[kpc]&[pc]&\\
    \hline
2MASS J00302193-7340258 &-0.019$\pm$0.067&-0.3&0.96&-1.00&9.97$^{+2.53}_{-2.48}$&-6844$^{+1739}_{-1706}$&Possible Be star\\
$[$MA93$]$ 332 &-0.681$\pm$0.267&-2.6&5.84&87.29&5.25$^{+3.91}_{-1.8}$&-3693$^{+2749}_{-1265}$&Possible Be star\\
2MASS J00520200-7300074 &0.109$\pm$0.067&1.6&0.94&-1.55&6.07$^{+1.98}_{-1.33}$&-4229$^{+1377}_{-929}$&Possible Be star\\
2MASS J05042350-6946395 &-0.288$\pm$0.154&-1.9&3.86&51.89&6.94$^{+2.73}_{-2.12}$&-3932$^{+1547}_{-1199}$&PN\\
2MASS J10230286-5420071 &-0.020$\pm$0.049&-0.4&1.00&0.12&9.55$^{+2.34}_{-1.94}$&415$^{+102}_{-84}$&D-type SySt\\
WRAY 15-678 &0.003$\pm$0.055&0.1&1.09&2.31&8.24$^{+2.35}_{-1.7}$&-159$^{+46}_{-33}$&D-type SySt\\
IRAS 11015-5850 &0.212$\pm$0.020&10.4&1.04&1.19&4.14$^{+0.54}_{-0.34}$&66$^{+9}_{-5}$&D'-type SySt\\
Mul 22 &0.090$\pm$0.047&1.9&1.02&0.59&7.33$^{+2.15}_{-1.75}$&1828$^{+536}_{-436}$&D-type SySt\\
Mul 20 &0.180$\pm$0.057&3.2&1.05&1.38&4.88$^{+2.08}_{-1.10}$&1154$^{+493}_{-261}$&D-type SySt\\
Mul 21 &0.163$\pm$0.114&1.4&1.12&2.51&5.39$^{+2.19}_{-1.38}$&-562$^{+229}_{-144}$&D-type SySt\\
2MASS J20385125+3111042 &0.622$\pm$0.014&44.3&0.98&-0.60&1.53$^{+0.03}_{-0.04}$&-165$^{+3}_{-4}$&D'-type SySt\\
2MASS J20581257+2920454 &1.500$\pm$0.074&20.3&1.01&0.33&0.66$^{+0.03}_{-0.03}$&-121$^{+6}_{-5}$&Polar CV\\
ATO J315.3668+45.9271 &0.269$\pm$0.079&3.4&1.63&15.76&3.49$^{+1.01}_{-0.59}$&-18$^{+5}_{-3}$&D-type SySt\\
IRAS 21136+5810 &0.099$\pm$0.017&5.7&1.03&0.91&7.69$^{+1.06}_{-0.71}$&882$^{+122}_{-81}$&D'-type SySt\\

\hline    
    \end{tabular}
    \tablefoot{$^a$Renormalised Unit Weight Error (RUWE). A value close to 1.0 suggests a good fit with a single-star model, whereas notably higher values may point to problems with the astrometric solution or the presence of a binary companion. $^b$Goodness-of-fit statistic for the astrometric solution. As per the \textit{Gaia} DR3 documentation, values of $\gtrsim$3 indicate a poor fit to the observed data.}
    \label{tab:GAIA_main}
\end{table}

\begin{table}[]
    
    \centering
      
     \setlength{\tabcolsep}{4.55pt}
     \renewcommand{\arraystretch}{1.1}
    \caption{Log of spectroscopic observations of our selected objects, and their proposed nature.}
    \begin{tabular}{lcccl}
    \hline\hline
Object & Exposure time & Julian Date & Obs. Site & Nature \\
    \hline
2MASS J00302193-7340258 &3 x 1200 s&2460934.7322&DSC&Possible Be star\\
$[$MA93$]$ 332 &3 x 1200 s&2460934.8007&DSC&Possible Be star\\
2MASS J00520200-7300074 &5 x 1200 s&2460935.8203&DSC&Possible Be star\\
2MASS J05042350-6946395&3 x 1200 s&2460756.5345&DSC&PN\\
2MASS J10230286-5420071&5 x 1200 s&2460756.6037&DSC&D-type SySt\\
WRAY 15-678&9 x 1200 s&2460756.7264&DSC&D-type SySt\\
IRAS 11015-5850&6 x 1200 s&2460765.7139&DSC&D'-type SySt\\
Mul 22&6 x 1200 s&2460567.5593&DSC&D-type SySt\\
Mul 20&2 x 1200 s&2460566.6062&DSC&D-type SySt\\
Mul 21&4 x 1200 s&2460566.5556&DSC&D-type SySt\\
2MASS J20385125+3111042&5 x 1200 s&2460886.4663&OHP&D'-type SySt\\
2MASS J20581257+2920454&8 x 1200 s&2460910.4236&Cornillon&Polar CV\\
ATO J315.3668+45.9271&5 x 1200 s&2459787.5084&OHP&D-type SySt\\
IRAS 21136+5810 &5 x 1200 s&2460887.4180&OHP&D'-type SySt\\

\hline
    \end{tabular}
    \label{tab:log_obs}
\end{table}

\begin{table}
    \centering
    \renewcommand{\arraystretch}{1.1}
    \caption{Infrared magnitudes from 2MASS survey of our selected objects.}
    \begin{tabular}[t]{lrrrrrl}
\hline\hline
Object &J [mag]&H [mag]&K [mag]&H-K [mag]&J-H [mag]&Nature\\
    \hline
2MASS J00302193-7340258 &17.10&15.94&15.73&0.21&1.16&Possible Be star\\
$[$MA93$]$ 332 &16.36&15.85&15.42&0.43&0.51&Possible Be star\\
2MASS J05042350-6946395 &15.99&15.75&14.38&1.37&0.24&PN\\
2MASS J10230286-5420071 &14.35&12.52&11.12&1.40&1.83&D-type SySt\\
WRAY 15-678 &13.38&11.17&9.26&1.92&2.21&D-type SySt\\
IRAS 11015-5850 &13.61&13.23&12.67&0.56&0.38&D'-type SySt\\
Mul 22 &15.59&14.90&13.42&1.48&0.69&D-type SySt\\
Mul 20 &15.33&14.45&12.84&1.61&0.88&D-type SySt\\
Mul 21 &16.46&15.60&14.09&1.51&0.86&D-type SySt\\
2MASS J20385125+3111042 &11.80&10.92&10.20&0.72&0.88&D'-type SySt\\
2MASS J20581257+2920454 &15.01&14.11&13.41&0.70&0.90&Polar CV\\
ATO J315.3668+45.9271 &9.39&7.83&6.75&1.08&1.56&D-type SySt\\
IRAS 21136+5810 &12.79&12.22&11.65&0.57&0.56&D'-type SySt\\
\hline    
    \end{tabular}
    \label{tab:2MASS_mag}
\end{table}

\begin{table}
    \centering
    \renewcommand{\arraystretch}{1.1}
    \caption{Infrared magnitudes from WISE satellite of our selected objects.}
    \begin{tabular}[]{lrrrrrrl}
\hline\hline
Object &W1&W2&W3&W4&W1-W2&W3-W4&Nature\\
&[mag]&[mag]&[mag]&[mag]&[mag]&[mag]&\\
    \hline
2MASS J00302193-7340258 &16.62&16.54&12.94&8.99&0.07&3.95&Possible Be star\\
2MASS J05042350-6946395 &13.04&12.02&7.50&4.51&1.03&2.99&PN\\
2MASS J10230286-5420071 &9.79&9.11&7.96&5.91&0.68&2.05&D-type SySt\\
WRAY 15-678 &6.84&5.17&2.43&0.74&1.67&1.69&D-type SySt\\
Mul 22 &11.49&10.46&6.58&4.66&1.03&1.92&D-type SySt\\
Mul 20 &11.08&10.04&6.31&4.36&1.04&1.95&D-type SySt\\
Mul 21 &11.99&10.87&8.14&6.44&1.13&1.70&D-type SySt\\
2MASS J20385125+3111042 &8.82&8.10&5.76&4.21&0.72&1.55&D'-type SySt\\
2MASS J20581257+2920454 &13.11&12.57&11.25&9.27&0.53&1.98&Polar CV\\
ATO J315.3668+45.9271 &5.23&4.31&2.08&1.23&0.92&0.85&D-type SySt\\
IRAS 21136+5810 &10.36&9.20&5.55&3.42&1.16&2.12&D'-type SySt\\
\hline    
    \end{tabular}
    \label{tab:WISE_mag}
\end{table}

\begin{table}
    \centering
    \renewcommand{\arraystretch}{1.1}
    \caption{Peak intensities of lines of interest derived from our observations (no extinction correction applied).}
    \begin{tabular}[t]{lrrrrrrrl}
\hline\hline
Object&[\ion{Ne}{iii}]&H$\gamma$&[\ion{O}{iii}]&\ion{He}{ii}&H$\beta$&[\ion{O}{iii}]&H$\alpha$&Nature\\
&$\lambda$3869&$\lambda$4340&$\lambda$4363&$\lambda$4686&$\lambda$4861&$\lambda$5007&$\lambda$6562&\\
    \hline
2MASS J05042350-6946395 &-&0.46&-&-&1&0.68&3.14&PN\\
2MASS J10230286-5420071 &0.77&0.44&0.36&-&1&2.96&5.18&D-type SySt\\
WRAY 15-678 &1.35&0.54&1.34&-&1&5.71&30.58&D-type SySt\\
IRAS 11015-5850 &1.02&0.51&1.00&0.52&1&3.07&5.62&D'-type SySt\\
Mul 22 &1.40&0.43&1.07&0.22&1&5.92&4.18&D-type SySt\\
Mul 20 &2.13&0.45&0.94&-&1&8.45&3.13&D-type SySt\\
Mul 21 &0.92&0.44&0.70&0.60&1&0.95&7.14&D-type SySt\\
2MASS J20385125+3111042 &-&0.46&0.83&0.50&1&1.05&4.20&D'-type SySt\\
2MASS J20581257+2920454 &-&1.14&-&0.73&1&-&0.62&Polar CV\\
ATO J315.3668+45.9271 &-&-&-&0.39&1&1.40&32.00&D-type SySt\\
IRAS 21136+5810 &-&0.50&0.78&0.47&1&1.69&5.60&D'-type SySt\\
\hline    
    \end{tabular}
    \label{tab:lines_intensities}
\end{table}

\begin{table}
    \centering
    \renewcommand{\arraystretch}{1.01}
    \caption{PNe and symbiotic stars set plotted in diagnostic diagrams in Fig.\ref{fig:Gutiérrez-Moreno-Ilkiewicz_diag}.}
    \begin{tabular}[t]{lll}
\hline\hline
Object&Nature$^a$&Spectrum\\
    \hline
Cn 1-1&Confirmed D-type SySt&\cite{1992secg.book.....A}\\
H 1-36&Confirmed D'-type SySt&\cite{1992secg.book.....A}\\
M 1-2&Confirmed D-type SySt&\cite{1992secg.book.....A}\\
Hen 2-101&Confirmed D'-type SySt&\cite{1992secg.book.....A}\\
Hen 2-104&Confirmed D'-type SySt&\cite{1992secg.book.....A}\\
Hen 2-171&Confirmed D'-type SySt&\cite{1992secg.book.....A}\\
JaSt 79&Confirmed D'-type SySt&HASH team$^b$\\
K 5-33&Confirmed D'-type SySt&HASH team$^b$\\
ShWi 2-5&Confirmed D-type SySt&HASH team$^b$\\
SaSt 1-1&Confirmed D-type SySt&HASH team$^c$\\
IC 4997\textdagger&True PN&HASH team$^c$\\
IC 5117&True PN&\cite{1992secg.book.....A}\\
M 2-29&Likely PN&\cite{1992secg.book.....A}\\
M 3-27&True PN&\cite{1992secg.book.....A}\\
NGC 6833&True PN&\cite{1992secg.book.....A}\\
PC 11&Possible PN&\cite{1992secg.book.....A}\\
Vy 2-2&True PN&\cite{1992secg.book.....A}\\
NGC 7027&True PN&\cite{1996ApJ...473..304K}\\
NGC 6886&True PN&\cite{2001ApJ...562..804K}\\
NGC 6790&True PN&\cite{2001ApJ...562..804K}\\
Hb 12&True PN&\cite{2003PASP..115...80K}\\
K 4-47&True PN&\cite{ 2010ApJ...724..748H}\\
NGC 6302&True PN&\cite{ 2010ApJ...724..748H}\\
NGC 6881&True PN&\cite{ 2010ApJ...724..748H}\\

\hline    
    \end{tabular}
    \tablefoot{$^a$Nature from NODSV for symbiotics and from HASH for PNe. $^b$Spectrum obtained at the Australian Astronomical Observatory with 3.9 m telescope. $^c$Spectrum obtained at South African Astronomical Observatory with 1.9 m telescope.\textdagger The [\ion{O}{iii}]$\lambda$5007 line is saturated; its intensity was estimated as three times that of [\ion{O}{iii}]~$\lambda4959$.}
    \label{tab:reference_sp}
\end{table}

\end{appendix}

\end{document}